\documentclass[preprint,11pt]{aastex}
\usepackage{epsfig,natbib,emulateapj5}
\shortauthors{Eracleous et al.}
\shorttitle{Three LINERs With {\it Chandra}}

\slugcomment{To Appear in the Astrophysical Journal}
\def\n404{NGC~404}
\def\n4736{NGC~4736}
\def\n4579{NGC~4579}
\def\liners{LINERs}
\def\liner{LINER}
\def\chandra{{\it Chandra}}
\def\hst{{\it HST}}
\def\asca{{\it ASCA}}
\def\rosat{{\it ROSAT}}

\begin{document}

\def\aj{\rm{AJ}}                   
\def\araa{\rm{ARA\&A}}             
\def\apj{\rm {ApJ}}                
\def\apjl{\rm{ApJ}}                
\def\apjs{\rm{ApJS}}               
\def\apss{\rm{Ap\&SS}}             
\def\aap{\rm{A\&A}}                
\def\aapr{\rm{A\&A~Rev.}}          
\def\aaps{\rm{A\&AS}}              
\def\mnras{\rm{MNRAS}}             
\def\nat{\rm{Nature}}              
\def\pasj{\rm{PASJ}}    	   
\def\procspie{\rm{Proc.~SPIE}}     

\title{Three LINERs Under the Chandra X-Ray Microscope}

\author{
Michael Eracleous\altaffilmark{1}, 
Joseph C. Shields\altaffilmark{2}, 
George Chartas\altaffilmark{1},
\\ and 
Edward C. Moran\altaffilmark{3,4}
}
\altaffiltext{1}{Department of Astronomy and Astrophysics, The
Pennsylvania State University, 525 Davey Lab, University Park, PA  16802
(mce@astro.psu.edu, chartas@astro.psu.edu)}

\altaffiltext{2}{Department of Physics and Astronomy, Ohio University,
251B Clippinger Labs, Athens OH 45701-2979,
(shields@helios.phy.ohiou.edu)}

\altaffiltext{3}{Department of Astronomy, University of California,
601 Campbell Hall, Berkeley, CA 94720-3411,
{(edhead@jester.berkeley.edu)}}

\altaffiltext{3}{\chandra\ Fellow}

\begin{abstract}
We use observations of three galaxies hosting low-ionization nuclear
emission-line regions (\liners; NGC~404, NGC~4736, and NGC~4579) with
the \chandra\ X-ray Observatory to study their power sources.  We find
very diverse properties within this small group: NGC~404 has an
X-ray-faint nucleus with a soft, thermal spectrum, NGC~4736 harbors a
plethora of discrete X-ray sources in and around its nucleus, and
NGC~4579 has a dominant nuclear point source embedded in a very
extended, diffuse nebulosity. From their multi-wavelength properties
we conclude the following about the power sources in these LINERs: the
nucleus of NGC~404 is the site of a weak, compact starburst, whose
X-ray emission is due to gas heated by stellar winds and supernovae,
NGC~4736 appears to be in a recent or aging starburst phase, where the
X-ray emission is dominated by a dense cluster of X-ray binaries, and
NGC~4579 is powered by accretion onto a supermassive black hole.  We
detect 39 discrete sources in NGC~4736 and 21 in NGC~4579, most with
$L_{\rm X}> 10^{37}~{\rm erg~s^{-1}}$. One source in the disk of
NGC~4579 appears to be an ultraluminous X-ray binary with $L_{\rm
X}(2-10~{\rm keV})=9\times 10^{39}~{\rm erg~s^{-1}}$, but it could
also be a background quasar. The most luminous discrete sources have
simple power-law spectra, which along with their luminosities suggest
that these are X-ray binaries accreting near or above the Eddington
rate for a neutron star.  By comparing the luminosity functions of
discrete X-ray sources in these and other galaxies we find a potential
connection between the age of the stellar population and the slope of
the cumulative X-ray source luminosity function: galaxies with
primarily old stellar populations have steeper luminosity functions
than starburst galaxies. We suggest that this difference results from
the contribution of high-mass X-ray binaries from the young stellar
population to the upper end of the luminosity function.
\end{abstract}

\keywords{galaxies: active -- galaxies: nuclei -- X-rays: galaxies --
X-rays: binaries}

\section{Introduction}

Low-ionization nuclear emission-line regions \citep[\liners;][]{h80}
are very common in nearby galaxies; they are found in at least 30\% of
all galaxies and in 50\% of nearby, early-type spirals. Their optical
spectra feature narrow emission lines with relative strengths
indicative of a low ionization state of the line-emitting gas.
Although a great deal of progress has been made recently in
characterizing the optical properties of \liners\
\citep[cf,][]{hfs_a97,hfs_b97,hfs_c97}, the nature of their power
source remains controversial. This is because a number of different
astrophysical processes can give rise to \liner-like optical spectra
\citep[see][for a review]{f96}. The possibilities include one or a
combination of the following processes.

\begin{enumerate}

\item
The original idea suggested by \citet{h80} was that the optical emission
lines originate in shocked gas in the nucleus of the host galaxy. In
their modern form, shock models can reproduce the relative strengths of
many of the emission lines quite well \citep[see][]{ds95,d_etal96}.

\item
A second possibility is that the lines originate in dense gas
photoionized by a hard X-ray continuum with a low ionization parameter
\citep{fn83,hs83}. This hypothesis has the important implication that
LINERs harbor a low-luminosity \citep[or intermittent;][]{elb95}
active galactic nucleus (hereafter AGN). Indeed, the discovery of
broad Balmer lines in the spectra of many \liners\ (some of them
transient or seen in polarized light) bolsters this view
\citep*{sbw93,bower_etal96,hfs_b97,barth_etal99,
s_etal00,h_etal00,barth_etal01,eh01}.

\item
Yet another possibility is that the dense gas emitting the optical
lines is illuminated by a compact starburst with hot stars. This was
originally considered by \citet{tm85}, who took extreme Wolf-Rayet
stars to be the dominant source of ionizing radiation. Since then this
scenario has been refined and expanded to include the effects of O
stars, a range of densities in the gas, and a population of
post-asymptotic giant branch stars
\citep{ft92,s92,binette_etal94,bs00}.

\end{enumerate}

Since \liners\ are so common, improving our understanding of their
power source is important and may have far-reaching implications.  The
frequency of \liners\ that {\it are} weak AGNs sets a lower limit to
the frequency of supermassive black holes in the centers of galaxies.
If the majority are really accretion-powered sources, they would be
the most common type of AGN in the present-day universe, representing
the faint end of the AGN luminosity function.  They could be the
remnants of once ferocious quasars, and as such their demographics
constrain the endpoint of quasar evolution.  Moreover, the observed
spectral energy distributions suggest that the accretion flows of weak
AGNs in \liners\ have a different structure than in Seyferts
\citep[e.g.,][]{l_etal96,h99}.  They may be described by advection
dominated accretion flows \citep[ADAFs;][]{ny94,ny95}, adiabatic
inflow-outflow solutions \citep[ADIOS;][]{bb99}, or related systems,
which are interesting in their own right for understanding the growth
of supermassive black holes and their radiative efficiency.  In those
\liners\ powered by compact starbursts, the very presence of the
associated young star clusters, and also their properties, constrain
formation scenarios of galaxy bulges \citep[e.g.,][]{b00}. More
specifically, the tidal influence of a compact star cluster in a
galaxy center can destroy a pre-existing galactic bar that led to the
formation of the bulge in the first place \citep{c99}.

In view of the importance of \liners\ as a class, we undertook a
study of their X-ray properties with the \chandra\ X-Ray Observatory,
in order to address some of the above issues. The results of previous
X-ray observations have left ambiguities since older instruments had
either high spatial resolution (e.g., the {\it Einstein} and \rosat\
HRIs) or spectroscopic capabilities (e.g., the {\it Einstein} IPC, the
\rosat\ PSPC, and the \asca\ SIS and GIS) {\it but not both}.  The
combination of high spatial resolution and hard X-ray response
afforded by \chandra\ makes it possible to resolve such ambiguities by
putting nearby \liners\ under the microscope and examining their power
sources in great detail. Accordingly, we have adopted the strategy of
obtaining long exposures of a small number of \liners.  Our program
thus complements shallower \chandra\ surveys of large, statistically
complete samples of nearby galaxies, which can address different types
of questions \citep[e.g.,][]{ho_etal01}.

In this paper we present the results of our detailed study of 3
\liners\ observed during \chandra\ Cycle~1. Included are two objects
observed as part of our own guest-observer program (NGC~4736 and
NGC~4579) and one object observed as part of a different program by a
different group (NGC~404; the data were obtained from the \chandra\
public archive; program 01700766, P.I. P. Lira). We present the
properties of the target galaxies in \S2 and give the details of the
observations and data reduction in \S3. After describing the X-ray
morphologies of the three galaxies in \S4, we embark on a detailed
analysis of their X-ray sources in \S5--7. We conclude with a
discussion of the results and their implications for the nature of
\liners\ as a class in \S8 and \S9.

\section{Targets and Their Properties}

Our targets are nearby, well-studied galaxies. They are listed in
Table~\ref{targtable} in order of increasing distance, along with their
basic properties. NGC~404 is a small elliptical galaxy while
NGC~4736 and NGC~4579 are nearly face-on spirals (the latter is a
barred spiral). All three galaxies were observed in the {\it Hubble
Space Telescope (HST)} ultraviolet snapshot survey of nearby galaxies
\citep{m_etal95,m_etal96} and were found to harbor unresolved, nuclear
point sources, hinting to the presence of an AGN. Later UV spectroscopy
of NGC~4579  and NGC~404 (also with the \hst) showed the former to be
a {\it bona fide} AGN with broad emission lines. The UV spectrum of
NGC~404, on the other hand, resembles very closely that of a starburst,
with prominent absorption lines from hot stars and no indication of an
AGN. This result also confounds the case for an AGN in NGC~4736,
since it shows that a point-like UV source could well be associated
with hot stars.  Narrow-band H$\alpha$ and [\ion{O}{3}] images of the
three galaxies, obtained from the ground and with the \hst\
\citep{p89,p_etal00}, resolve their nuclear emission-line regions
down to scales of less than 10~pc and show complex filamentary
structure. It is noteworthy that the H$\alpha$-emitting region of
NGC~4579 is easily resolved from the ground; it extends along the
direction of the bar of the host galaxy, with a projected size of
approximately 3~kpc. The nucleus of NGC~4736 is surrounded by a
star-forming ring of diameter 2~kpc, seen very clearly in the
H$\alpha$ images of \citet{p89}, as well as in the radio continuum
image of \cite{th94}.

The target galaxies have been observed several times in the X-ray
band. The two spirals are fairly bright X-ray sources in both soft and
hard X-rays. Their \asca\ spectra \citep*{t_etal98,rwo99} can be
described by a composite model consisting of a hard power law with a
Seyfert-like photon index of $\sim 1.7$ plus thermal plasma emission
with a characteristic temperature in the range 0.1--1~keV, just like
the spectra of most \liners\ \citep{p_etal99}. The observed
luminosities are $L_{\rm X} (2-10~{\rm keV})=1.5 \times 10^{41}~{\rm
erg~s^{-1}}$ for NGC~4579 and $L_{\rm X} (2-10~{\rm keV})=3\times
10^{39}~{\rm erg~s^{-1}}$ for NGC~4736.  Moreover, the \rosat\ HRI
image of NGC~4736 shows remarkable extended structure in the form of a
halo surrounding the nucleus \citep{h_etal01}, which is comparable in
size with the star-forming ring seen in the optical emission-line and
radio continuum images. The nucleus of NGC~4736 was even resolved by
the \rosat\ PSPC, which measured a soft X-ray luminosity of $L_{\rm X}
(0.1-2~{\rm keV})=3\times 10^{39}~{\rm erg~s^{-1}}$ \citep{cui97}. In
marked contrast, NGC~404 is extremely faint even though it is the
nearest of the three targets. It was barely detected in soft X-rays by
the \rosat\ HRI with a luminosity of $L_{\rm X} (0.1-2.4~{\rm
keV})\approx 5\times 10^{37}~{\rm erg~s^{-1}}$ \citep{kbc99}, while an
\asca\ observation yielded only an upper limit on the hard X-ray
luminosity of $L_{\rm X} (2-10~{\rm keV})< 5\times 10^{37}~{\rm
erg~s^{-1}}$ \citep{t_etal_a00}.

\section{Observations and Data Screening}

The three target galaxies were observed with the Advanced CCD Imaging
Spectrometer (ACIS; Garmire et al. 2001, in preparation) on \chandra\
in 1999 December and 2000 May. The instrument was operated in faint
mode. A log of observations, which includes the observation date and
exposure time for each target, is given in Table~\ref{obstable}. Their
nuclei were centered on the aimpoint of the S3 CCD, which had not
suffered significant radiation damage during the early stages of the
mission. In the observations of NGC~4736 and NGC~4579 only one quarter
of the S3 CCD was read out; all other CCDs were turned off in order to
reduce the frame time to 0.84~s, and thus to reduce the probability of
photon pileup in the central pixels of bright point sources. Even with
this measure, however, the bright point source in the nucleus of
NGC~4579 still suffered from pileup, which complicates the spectral
analysis (the circumnuclear regions of NGC~4579 were not affected). In
the observation of NGC~404, 6 CCDs were turned on (4 from the ACIS-S
array and 2 from the ACIS-I array), resulting in a frame time of
3.2~s. We consider here only the part of the image included on the S3
chip since this encompasses the entire optical galaxy.

The initial data screening was carried out with the \verb+CIAO+
software package, developed and distributed by the \chandra\ X-Ray
Center (CXC). Since this software was undergoing revisions over the
course of our analysis of the data, we processed the data using both
versions 1.0 and 2.0 to check for consistency. Some of the later steps
in the data analysis were also checked with version 2.1 of \verb+CIAO+.
In summary, the screening consisted of selection of events with
grades 0, 2, 3, 4, and 6 and exclusion of events recorded during times
of poor aspect solution and background flares. In the process we also
removed the 0\farcs 5 pixel randomization introduced 
during the initial data processing, which amounts to an unnecessary
smearing of the image. After screening, we produced images and spectra
from the photon event lists using the  \verb+CIAO+ tools and the
\verb+TARA+ software package \citep{broos_etal00}. Below we discuss the
morphology of the target galaxies as seen in the X-ray images and
compare with what is seen in other bands.

\begin{figure}
\plotone{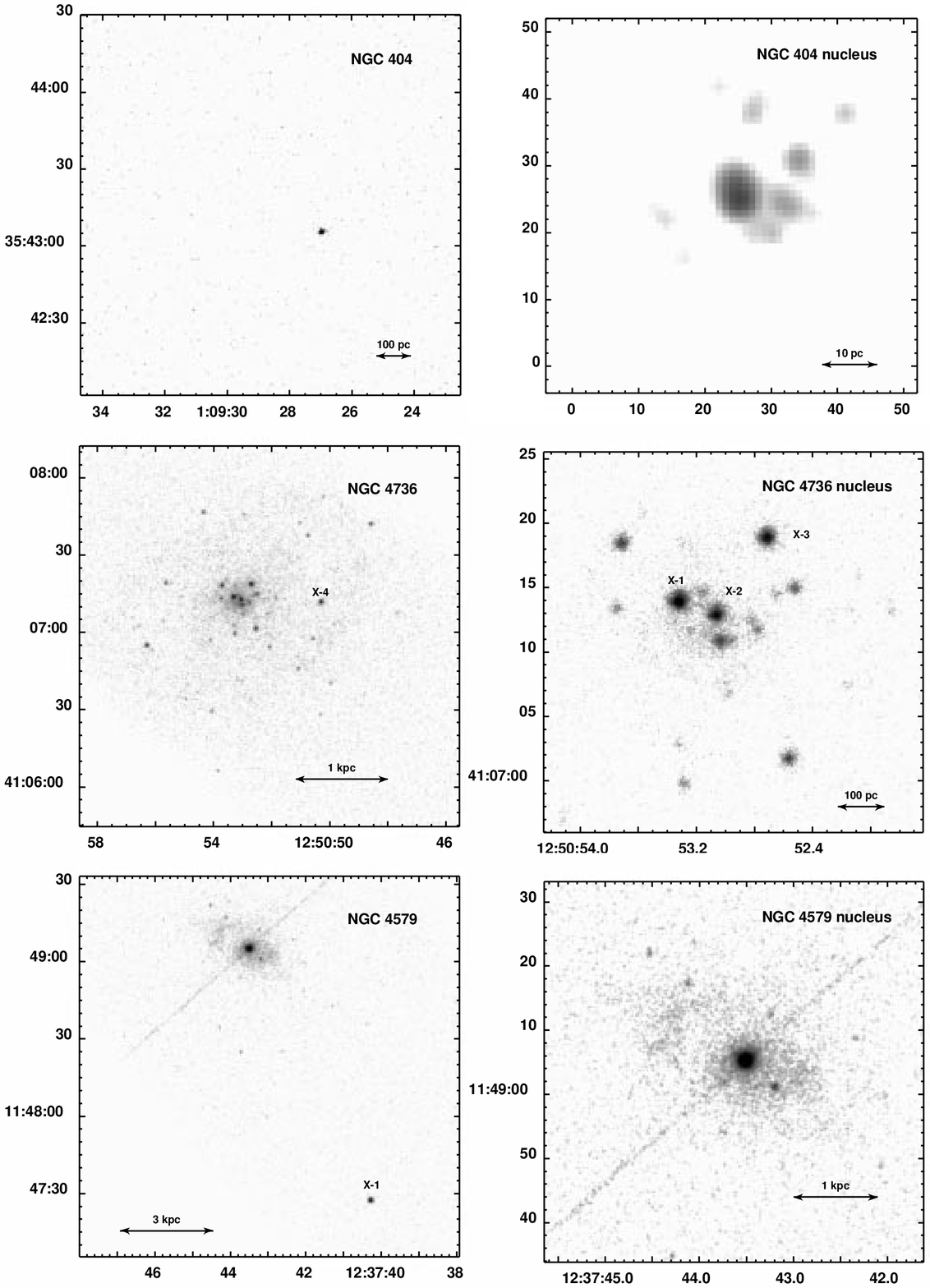}
\bigskip
\caption{\label{imgmont} Montage of \chandra\ images of the target
galaxies. For each galaxy we show an image encompassing all the
interesting features in the field of view as well as an enlarged view
of the nucleus. All but one of the frames are labeled with the right
ascension along the horizontal axis and the declination along the
vertical axis (the coordinate epoch is J2000, north is up, and east is
to the left). The image of the nucleus of NGC~404 has been deconvolved
using the Lucy-Richardson algorithm; its frame is calibrated in
pixels, where each pixel corresponds to $0.\!\!^{\prime\prime}15$.
The discrete ``blobs'' around the nucleus of NGC~404 are not
discernible in the original, unprocessed image. A horizontal bar in
each image gives the linear scale at the distance of that
galaxy. Bright X-ray sources, referred to in the text and in the
tables, are labeled on the images.}
\end{figure}

\section{Morphology}

Figure~\ref{imgmont} shows a montage of the 0.5--10~keV \chandra\
images of the target galaxies. For each galaxy we present an image
encompassing most of the ACIS-S3 field of view as well as a zoomed
display of the nucleus. These images show exquisite detail, which we
describe below in a separate paragraph for each galaxy. In our
comparison with images taken at other wavelengths we have assumed that
the ``nucleus'' of each galaxy has the same coordinates at all
wavelengths. This is certainly true within the {\it a priori} relative
astrometric uncertainties between images, which are approximately
$1^{\prime\prime}\!\!.5$ \citep[dominated by the \chandra\ aspect
solution;][]{annh01}. Because of the small field of view and
bandpasses of the \chandra\ and \hst\ images that we are comparing, we
are not able to find matching sources that can serve to improve the
relative registration of the images. We have also searched the
USNO-A2.0 catalog \citep{monet96} for optical counterparts to the
detected X-ray sources, but found none.

\begin{itemize}

\item{\it NGC 404. --} The only significant X-ray emission within the
optical extent of NGC~404 comes from a compact region associated with
the nucleus.  Figure~\ref{imgmont} (top right panel) shows an image of
the nucleus deconvolved via the Lucy-Richardson algorithm. There is
clearly-resolved extended emission to the west and southwest of the
nucleus, which after deconvolution takes the form of discrete
``blobs.''  The extended X-ray emission coincides with the extended UV
and line emission seen in the \hst\ images of NGC~404, although there
is also some low-level X-ray emission as far out as 10$\arcsec$ from
the center. Most of the flux is concentrated within 1$\arcsec$ from
the nucleus, which corresponds to a linear scale of 12~pc. The X-ray
spectrum is extremely soft, as we show in later sections. This fact
along with the shell-like structure of the X-ray and emission-line
images suggests that the X-ray source is hot gas, very likely
associated with a superbubble blown out by a compact starburst in the
nucleus of NGC~404. Contrary to the two spiral galaxies in our
collection and other galaxies observed with \chandra\
\citep*[including both spirals and
ellipticals;][]{sib01,bsi01,w_etal00,m_etal00,alm01,lira01,fzm01} we
do not find any discrete X-ray sources away from the nucleus of
NGC~404, down to a limiting luminosity of $1\times 10^{36}~{\rm
erg~s^{-1}}$.  \citep[Although there are many other sources in the
field of view, none of them fall within the effective radius of the
galaxy reported by][]{f_etal89}.

\item{\it NGC 4736. --} 
The image of NGC~4736 shows a plethora of discrete X-ray sources, with
a concentration around the nucleus of the galaxy. The collective
emission from these sources, which are most likely X-ray binaries,
makes up the extended, unresolved ``halo'' seen in the \rosat\ HRI
image, and accounts for most of the X-ray flux from this galaxy.  In
the innermost $400\times 400$~pc of the galaxy we find a very dense
cluster of 10 discrete sources embedded in a diffuse but clumpy
halo. The four brightest sources (3 of which are in the central
cluster) are labeled as X-1 through X-4 on the image (in order of
decreasing X-ray brightness). We find that within the combined
astrometric uncertainties of \chandra\ and the \hst, the southern of
the two UV point sources in the image of \citet{m_etal95} coincides
with source X-2, while the northern of the two UV sources appears to
have no X-ray counterpart. The detection of large numbers of discrete
X-ray sources both in NGC~4736 and in NGC~4579 (below) is undoubtedly
aided by the fact that the two galaxies are oriented face-on and have
a low Galactic absorption column. The spectra of the discrete X-ray
sources are fairly hard and their luminosities reach values well over
$10^{38}~{\rm erg~s}^{-1}$. The spectrum of the diffuse emission, on
the other hand, is softer with a total luminosity comparable to that
of one of the brightest discrete sources. We describe the spectral
properties of the discrete sources and the diffuse emission in more
detail in later sections. The large concentration of discrete X-ray
sources may be associated with a recent episode of star formation,
which is still ongoing to some degree in the circumnuclear
star-forming ring. We discuss this possibility further in \S8.2.

\item{\it NGC 4579. --} The X-ray emission from NGC~4579 is dominated
by an unresolved source coincident with the nucleus, which contributes
90\% of the 2--10~keV X-ray flux. We identify this source with the
AGN, which is the source of the broad optical and UV emission lines
\citep{barth_etal01, m_etal98}. The AGN is embedded in an extended,
diffuse X-ray halo, which is elliptical in shape. The halo major axis
is $\sim 2$~kpc in length, aligned approximately parallel to the bar
of the host galaxy. The X-ray halo appears to coincide exactly with
the H$\alpha$ emission region seen in the narrow-band image of
\citet{p89}, while the much more compact H$\alpha$-emitting region
seen in the \hst\ images of \citet{p_etal00} corresponds to the
innermost, brightest part of the X-ray halo, immediately surrounding
the AGN. The image of NGC~4579 also shows about 20 discrete X-ray
sources scattered throughout the disk of the host galaxy. These are
presumably X-ray binaries, just as in NGC~4736. One of these sources,
located at a projected distance of 8.7~kpc from the AGN at a position
angle of 206$^{\circ}$, is an ultraluminous X-ray binary (UXB), with
$L_{\rm X} (2-10~{\rm keV})= 9\times 10^{39}~{\rm erg~s^{-1}}$.  It is
unresolved by \chandra, which sets an upper limit of 120~pc to its
diameter.

\end{itemize}

In the next section we present fits to the spectra of bright sources
found in the three galaxies, while in later sections we study their
variability and population properties.

\section{Model Fits to the Spectra of Bright Sources}

\subsection{Extraction of Spectra}

To study the spectral properties of X-ray sources in the target
galaxies, we extracted their spectra using \verb+CIAO+. In particular,
we extracted spectra of (a) the nucleus of NGC~404 and patches of
diffuse emission in the nuclei of NGC~4736 and NGC~4579, (b) the AGN
in NGC~4579, and (c) the 4 brightest discrete sources (X-1 through
X-4) in NGC~4736 and the extranuclear UXB (i.e. X-1) in NGC~4579. In
all cases the spectra were extracted from a circular region centered
on the source. The radius of this region was generally 4.1~pixels
(corresponding to 2\arcsec) for discrete sources, which encompasses
more than 95\% of the flux from a point source. In the case of the AGN
in NGC~4579, which is quite bright and has a relatively hard spectrum,
we enlarged the radius of the extraction region to 6.2~pixels (or
3\arcsec) to include the wings of the point-spread function (PSF). In the
case of the extended emission in NGC~4736, which underlies a crowded
field of discrete sources, we extracted a spectrum from a region that
falls SE of source X-2 and SW of source X-1. This region has a radius
equal to a discrete source extraction radius and is free of discrete
sources \footnote{The absence of discrete sources was verified both by
visual inspection and by the systematic search described in \S7}.  In
the case of the extended emission from NGC~4579, we extracted a
spectrum from an annulus centered on the AGN, with an outer radius of
20 pixels (or $9^{\prime\prime}\!\!.8$). This annulus excludes the AGN
and has holes to exclude discrete sources that fall within it, as well
as the ``readout streak''. Its outer radius is small enough not to
cross to a region of the CCD read out by a different amplifier. The
response matrix and effective area curves corresponding to each source
were computed automatically by \verb+CIAO+ after the spectrum was
extracted.

A spectrum of the background for each observation was extracted from a
region of radius 50 pixels (or $24^{\prime\prime}\!\!.5$) centered in
a part of the field of view that appears free of discrete sources or
diffuse emission. The 0.5--10~keV background count rates in NGC~404,
NGC~4736, and NGC~4579 are, respectively, $6.4\times 10^{-7},
2.5\times 10^{-6},~{\rm and}~1.3\times 10^{-6}~{\rm
s^{-1}~pixel^{-1}}$, implying a total of 0.9, 6.1, and 2.3 background
counts per point source, respectively (assuming a radius of
2\arcsec). This background determination, however, applies to the
particle and diffuse sky background only. The effective background for
the discrete sources of interest can be considerably higher (and with
a different spectrum) because these sources are embedded in diffuse
emission from the host galaxy. A serious, additional complication
arises from the fact that the diffuse emission is patchy and uneven,
which prohibits us from using its spectrum, normalized by a geometric
factor, as an effective background spectrum. To overcome this
difficulty, we include the effective background in the model that we
fit to discrete source spectra. More specifically, we fit the spectrum
of the diffuse emission first and the model we derive for it is added
to the spectral model for each source. All the parameters of the
effective background model are held fixed, except for a normalization
factor.

The spectra of the diffuse sources in NGC~4736 and NGC~4579 could also
be contaminated by photons from the wings of the PSF of nearby, bright
discrete sources. These contaminating photons have a relatively hard
spectrum since the PSF is more extended at higher energies. To
eliminate this contamination we simulated the images of point sources
that could affect the spectra of the diffuse sources in the two
galaxies using the \verb+MARX+ tool. From the simulated images we
extracted the spectra of contaminating photons and subtracted them
from the observed spectra of diffuse sources.

Spectral fits were carried out using \verb+XSPEC+ v.11.0.1
\citep{a96}.  The spectra were truncated at the low end at 0.5~keV and
at the high end at energies generally between 2 and 7~keV, depending
on the signal-to-noise ratio ($S/N$). The spectral models described
below were modified by interstellar photoelectric absorption with
cross-sections as given by \cite{mm83}. The absorbing columns were
constrained to be at least as large as the Galactic columns given in
Table~\ref{targtable}. The abundance ratios used in thermal plasma
emission models are those of \citet{ag89}. The error bars given on
model parameters correspond to 90\% confidence intervals (for 2
interesting parameters), unless specified otherwise.

\subsection{Spectra of Diffuse Sources}

\begin{figure}
\plotone{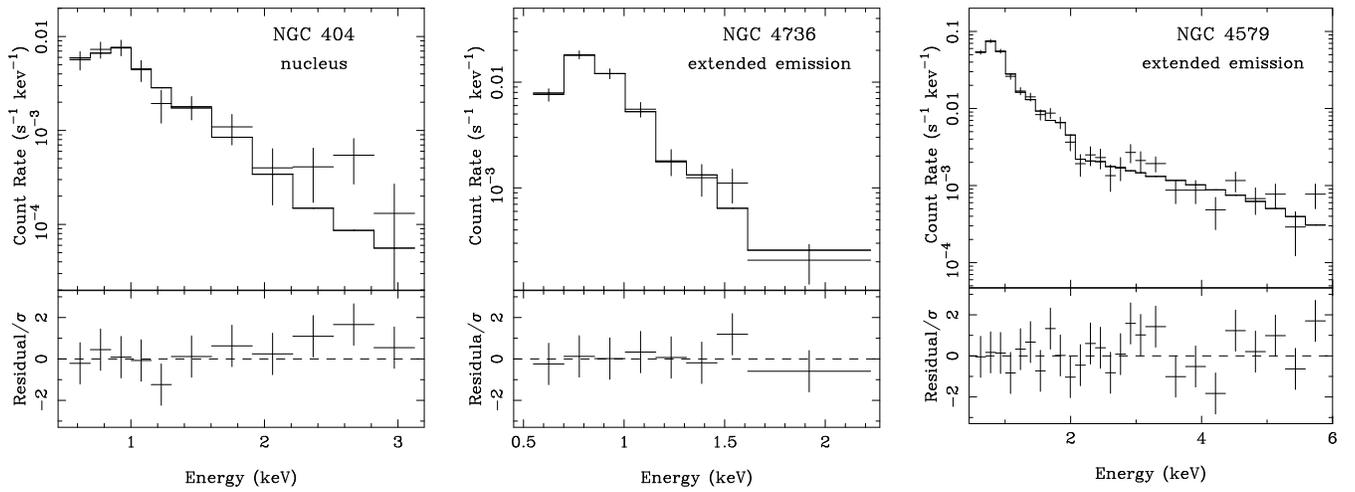}
\caption{\label{fuzzspec}
Spectra of the thermal plasma regions in the three target galaxies,
the nucleus of NGC~404 and the extended, circumnuclear emission in
NGC~4736 and NGC~4579. For each object we show the observed spectrum
with the best-fitting model superposed (upper panel) and the post-fit
residuals scaled by the error bar (lower panel). The models consist of
linear combinations of thermal plasma components, with abundances left
as free parameters (1 component in NGC~404 and NGC~4736, and 3
components in NGC~4579).  The parameters describing these models are
summarized in Table~\ref{fuzztable}.  }
\end{figure}

To model the spectrum of the diffuse emission in NGC~4736 and
NGC~4579, as well as the spectrum of the nucleus of NGC~404, we tried
combinations involving thermal plasma emission components
\citep*{mewe85,mewe86,kaastra92,liedahl95} with and without a power
law. Single-temperature plasma models provide acceptable fits to the
spectra of NGC~404 and NGC~4736, while a 3-temperature model is
required for NGC~4579. The parameters of the best-fitting models are
summarized in Table~\ref{fuzztable}, and the spectra themselves are
shown in Figure~\ref{fuzzspec} with models superposed. The
temperatures for the first two objects are around 0.6~keV, similar to
what was inferred from previous \asca\ observations of many \liners\
\citep[e.g.,][]{p_etal99}. In NGC~4579 one of the thermal emission
components also has a temperature of 0.6~keV, while the other two
temperatures bracket this value. It is likely that there is a range of
temperatures in all cases, although this is only discernible in
NGC~4579 where the $S/N$ is highest. It is also noteworthy that the
heavy-element abundances are considerably below their Solar values in
NGC~404. In NGC~4736 and NGC~4579 the abundances are consistent with
the Solar values. Keeping the abundance value in the latter object
fixed to Solar yields slightly different temperatures than if the
abundances were free to vary, as we indicate in Table~\ref{fuzztable}.

Models consisting of thermal plasma emission plus a power-law can also
yield acceptable fits, although such models are not preferable to the
ones described above in a strict statistical sense. In the case of
NGC~404, the power-law component has a photon index of
$2.5^{+1.0}_{-0.6}$, a 0.5--2~keV luminosity of $7\times 10^{36}~{\rm
erg~s^{-1}}$ and an extrapolated 2--10~keV luminosity of $5\times
10^{36}~{\rm erg~s^{-1}}$, equivalent to that of a single X-ray
binary. In NGC~4579, the hottest thermal component can be substituted
with a power-law component of photon index $1.5^{+0.4}_{-0.5}$ (for
$Z=Z_{\odot}$).  More importantly, however, the addition of a
power-law to the model for NGC~404 allows us to fit the spectrum
adequately by keeping the abundances fixed to their Solar values. The
fundamental reason for this is that the emission-line complexes in the
observed spectrum appear to be weaker than what thermal plasma models
predict. Thus the addition of the featureless power law dilutes the
line equivalent widths, and allows for acceptable fits with Solar
metal abundances. With pure thermal plasma models, on the other hand,
we can only achieve acceptable fits with metal abundances
significantly less than Solar.

There are two ways of interpreting the dilution of the thermal plasma
lines of NGC~404. One possibility is that the abundances are indeed
lower that Solar, while the other possibility is that there is a hard
X-ray source with a featureless power-law spectrum embedded in its
nucleus. As we discuss in more detail in {\S}8.1, the nucleus of
NGC~404 is best interpreted as a mild starburst, which most likely
hosts newly formed X-ray binaries. Therefore, we favor the latter
explanation of the apparent weakness of the thermal plasma emission
lines.

\subsection{Spectra of Discrete Sources}

We fitted the spectra of bright discrete sources with a variety of
models inspired by the results of studies of X-ray binaries in the
Milky Way and the Magellanic Clouds. The X-ray spectra of Galactic
X-ray binaries are typically described by composite models consisting
of either a multi-temperature accretion disk spectrum plus a
Comptonized blackbody spectrum \citep[motivated by {\it TENMA}
observations;][]{mitsuda84,mitsuda89} or by a single-temperature
blackbody spectrum plus unsaturated Comptonization \citep*[motivated
by {\it EXOSAT} observations;][]{wsp88}. More recent observations with
{\it BeppoSAX} and {\it RXTE} confirm these findings and show that
these models can sometimes be indistinguishable, even with high-$S/N$
spectra \citep[see review by][and references therein]{barret01}. The
above models apply to objects in their ``high'' luminosity state
($L_{\rm X} \sim 10^{37}-10^{38}~{\rm erg~s}^{-1}$). As the luminosity
approaches and exceeds $10^{38}~{\rm erg~s}^{-1}$, the spectral shape
resembles a simple power law very closely. Accordingly, we
experimented with combinations of soft and hard spectral components,
with the soft component taken to be either a black body or a
multi-color disk and the hard component taken to be a Comptonized
blackbody spectrum \citep{tit94} or a simple power law. We also
experimented with single-component models made up of either a simple
power law or a Comptonized blackbody spectrum.

\begin{figure}
\plotone{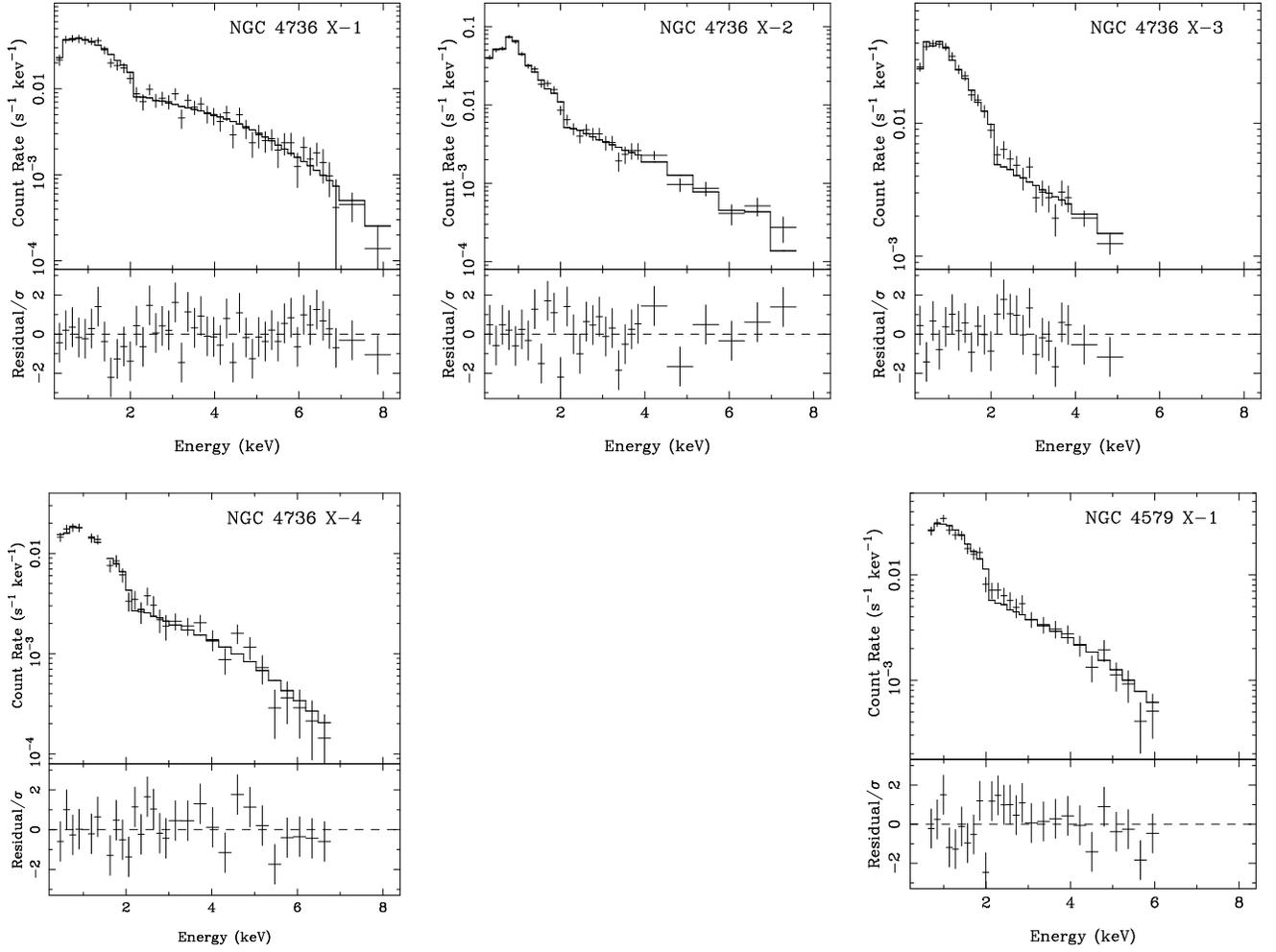}
\caption{\label{xrbspec}
Spectra of the brightest X-ray binaries in NGC~4736 and NGC~4579. The
objects are labeled according to the conventions defined in
Tables~\ref{sourcetable4736} and \ref{sourcetable4579}. For each object 
we show the observed spectrum with the best-fitting model superposed
(upper panel) and the post-fit residuals scaled by the error bar (lower
panel). The parameters describing the best-fitting models are summarized
in Table~\ref{powertable}.}
\end{figure}

We find that all spectra can be adequately described by a simple power
law (resulting in $\chi^2_{\nu}\leq 1.1$), with photon indices in the
range 1.1--1.8 and 2--10~keV luminosities between $4\times 10^{38}$ and 
$9\times 10^{39}~{\rm ~erg~s}^{-1}$ (the most luminous source is, of
course, the UXB in NGC~4579). The parameters of the individual power-law
models are given in Table~\ref{powertable}, while the observed spectra
with models superposed are shown in Figure~\ref{xrbspec}. In two cases we
measure absorbing columns that are almost an order of magnitude higher
than the Galactic column, suggesting significant absorption in the
vicinity of the source. These large columns should be regarded with
caution, however, because they do not appear to be so large in all the
models that we experimented with. The fact that simple power-law models
provide a good description of the observed spectra is consistent with
the very high luminosities of these objects, just as in Galactic X-ray
binaries. Most of the other spectral models described above fit the
observed spectra equally well. However, their larger number of free
parameters makes them less appealing in a strict, statistical sense.
Moreover, degeneracy between the model parameters and the limited $S/N$
and spectral coverage prevent us from placing meaningful constraints on
the parameter values. We note, nevertheless, that the fits of simple
Comptonization models yield parameters comparable to those found in
Galactic X-ray binaries, namely, seed blackbody temperatures between
0.2 and 1~keV, Comptonizing plasma temperatures between 2 and 5~keV, and
optical depths between 2 and 6. The UXB in NGC~4579 is a notable
exception in that a two-component model consisting of a multi-color
disk ($kT_{\rm inner}=0.4\pm 0.2$~keV) and Comptonized blackbody
($kT_{\rm bb}=0.5^{+0.6}_{-0.2}$~keV, $kT_{\rm compt}<200$~keV,
$\tau<1$) provides a somewhat better fit than a simple power-law model
($\chi^2$ decreases by 4.71 at the expense of adding 3 more free
parameters, which is significant only at the 73\% level). 

\subsection{The spectrum of the AGN in NGC~4579}

\begin{figure}
\plotone{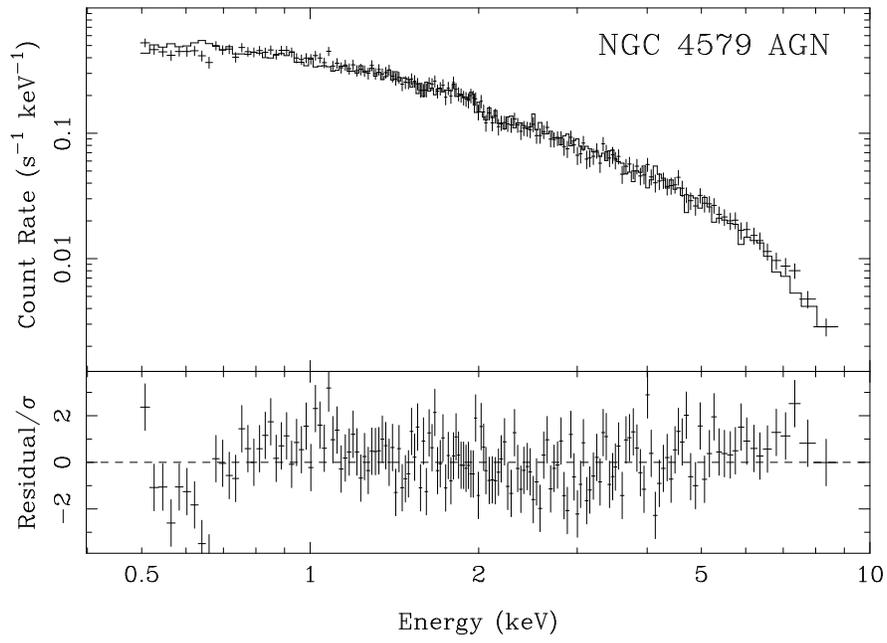}
\caption{\label{agnspec}
Spectrum of the AGN in NGC~4579 with the best-fitting model
superposed.  The model consists of a simple power law modified by
photoelectric absorption in the Galaxy, as described in \S5.4.  The
lower panel shows the residuals scaled by the error bar at each bin.
The large residuals around 0.6~keV may be the result of poor calibration
of the S3 CCD. Due to calibration uncertainties, we are not able to 
assess the reality of this feature.}
\end{figure}

Because of its high (observed) count rate of $0.719\pm 0.005~{\rm
s}^{-1}$, the spectrum of the AGN in NGC~4579 is severely distorted by
the effects of pileup. In practice, the main consequence of pileup is
to make the spectrum appear harder than it really is. This is a result
of two effects: (a) two or more lower-energy photons arriving in the
same pixel within one frame time are misidentified as single,
higher-energy photons, and (b) two or more photons arriving in the
same or adjacent pixels in one frame time are completely missed
because they produce non-standard event grades (this affects soft
photons more than hard photons because of their greater numbers).
Pileup cannot yet be incorporated into the detector response matrices
and effective area curves, thus piled-up spectra must be modeled using
special techniques. To this end, we have used the simulator-based,
forward-fitting tool \verb+LYNX+, developed by the ACIS instrument
team \citep{chartas00}. This tool simulates the distribution of counts
per energy channel for an assumed input spectrum and iteratively
refines the parameters of the model until the simulated count
distribution matches the observed one.  The simulation takes into
account photon trajectories (with the help of the \verb+MARX+ tool)
and pileup.  Uncertainties in the model parameters are computed
statistically from the distribution of their values over a large
number of realizations of the model. Using the above method, we were
able to fit the AGN with a simple power-law model, with the absorbing
column density fixed at the Galactic value. The resulting photon index
is $\Gamma=1.88\pm0.03$ and the 2--10~keV flux and luminosity (with
pileup taken into account) are, respectively, $F_{\rm X} = 5.2\times
10^{-12}~{\rm erg~cm^{-2}~s^{-1}}$ and $L_{\rm X} = 1.7\times
10^{41}~{\rm erg~s^{-1}}$. This model yields a reduced $\chi^2$ value
of 0.773 for 646 degrees of freedom. The observed spectrum is shown in
Figure~\ref{agnspec} with the best-fitting model superposed for
comparison.  We note for reference that fitting the observed spectrum
without taking pileup into account yields $\Gamma=1.20$ and $F_{\rm X}
= 6.3\times 10^{-12}~{\rm erg~cm^{-2}~s^{-1}}$. The degree of pileup
can be quantified by the ``pileup fraction'', which is defined as the
ratio of the {\it excess} counts in a selected energy window to the
true counts in that window. In this case we estimate the pileup
fraction in the 0.5--10~keV band to be $f_{\rm p}(0.5-10~{\rm
keV})=-0.51$ and the pileup fraction in the 6--7~keV window to be
$f_{\rm p}(6-7~{\rm keV})=0.33$. In other words, the 0.5--10~keV count
rate is 1.51 times {\it less} than the true value and the and the
6--7~keV count rate is 1.33 times {\it more} than the true value.

To check the results of the pileup simulation, we extracted spectra of
the AGN from an annulus that includes only the wings of the PSF and
excludes the core where the pileup occurs. We took the outer radius of
the annulus to be 6\arcsec\ and we experimented with inner radii in
the range $0.\!\!^{\prime\prime}5 - 1.\!\!^{\prime\prime} 2$. To fit
these spectra we modified the effective area curves by an
energy-dependent scale factor that takes into account the fact that
the size of the PSF depends on energy (this factor was determined with
the help of synthetic PSFs simulated with the \verb+MARX+ tool). We
found that these spectra became progressively steeper as the inner
radius of the annulus was increased, i.e., as more of the core of the
PSF was excluded. For inner radii greater than 1\arcsec\ the photon
index settled to a value consistent with the \verb+LYNX+ result.

\begin{figure}
\plotone{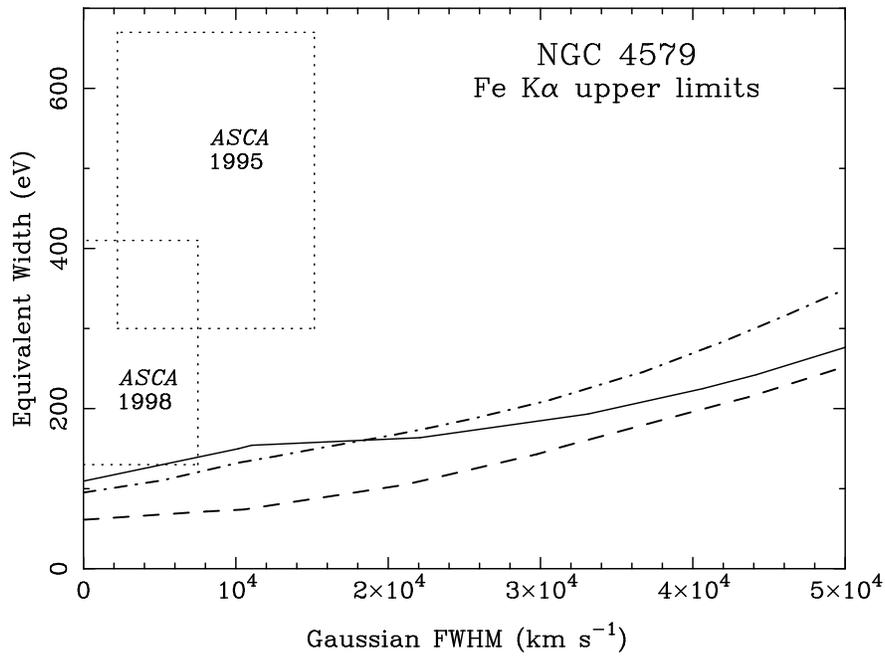}
\caption{\label{linecont}
Confidence contours showing the 99\% upper limits on the equivalent
width of the Fe~K$\alpha$ line in the spectrum of the AGN in NGC~4579
as a function of its FWHM. The equivalent width scale has been
corrected for the effect of pileup by multiplying all values by
$1+f_{\rm p}$, where $f_{\rm p}$ is the pileup fraction in the
6--7~keV band (see \S5.4 of the test for a detailed discussion). The
three curves refer to different assumed line energies; solid curve:
6.4~keV, dashed curve: 6.7~keV, dot-dashed curve: 6.9~keV. The error
boxes (dotted lines corresponding to the 90\% confidence regions) show
the measurements from {\it ASCA} observations
\citep{t_etal98,t_etal_b00} with dates as marked.}
\end{figure}

We also searched for the Fe~K$\alpha$ line detected in previous {\it
ASCA} observations in 1995 and 1998 by \citet{t_etal98,t_etal_b00},
but we did not detect it. The method we adopted was to isolate the
spectrum in the 4--8~keV range and fit it with a model consisting of a
power-law continuum plus Gaussian emission lines at 6.4, 6.7, and
6.9~keV.  The power-law model for the continuum was only a
parameterization, with no particular physical significance. The effect
of pileup on this portion of the spectrum is to raise the continuum
level and reduce the photon index by transforming pairs of
lower-energy photons into {\it apparent} high-energy photons.  Thus,
it effectively ``dilutes'' the observed equivalent width of any lines
in this window. The amount of dilution of lines between 6 and 7~keV is
the pileup fraction estimated above, $f_{\rm p}(6-7~{\rm
keV})=0.33$. Thus all equivalent widths or upper limits determined by
our procedure can be rectified by multiplying by $1+f_{\rm
p}$. Accordingly, we plot in Figure~\ref{linecont} the 99\%-confidence
upper limits to the line equivalent width as a function of the assumed
velocity width (FWHM). We also mark the measured equivalent widths and
FWHM from the older, {\it ASCA} observations, which appear to be
inconsistent with our upper limits. It is noteworthy that the photon
index and 2--10~keV flux that we measure, also differ somewhat from
what was measured by {\it ASCA} in 1995 ($\Gamma=1.72\pm 0.02$,
$F_{\rm X} = 4.3\times 10^{-12}~{\rm erg~cm^{-2}~s^{-1}}$), but are
consistent with the {\it ASCA} measurements from 1998 ($\Gamma=1.81\pm
0.06$, $F_{\rm X} = 5.3\times 10^{-12}~{\rm erg~cm^{-2}~s^{-1}}$).
Although it is conceivable that the Fe~K$\alpha$ line intensity has
varied between the the \chandra\ and \asca\ observations, we regard
this as a rather unlikely explanation since the luminosity and
spectral index appear to be the same now as in the 1998 \asca\
observation.

A more plausible and likely explanation for the difference between the
\asca\ and \chandra\ measurements is the fact that the \asca\ spectra
were extracted from a much larger region than the \chandra\ spectra
and include emission from diffuse thermal plasma as well as many
discrete sources.  To put this difference in perspective, we note that
for the AGN in NGC~4579 the area of the \asca\ SIS extraction region
($r=4^{\prime}$) is 6,400 times larger than the area of a \chandra\
extraction region ($r=3^{\prime\prime}$). Moreover, the radius of the
\asca\ SIS extraction region is twice as large as the radius of the
entire optical galaxy.  Thus any emission from hot plasma in the disk
of NGC~4579 (analogous to the ``Galactic Ridge'' emission) is included
in the \asca\ spectrum and contributes to the strength of the
Fe~K$\alpha$ line. The fact that the line energy was 6.7~keV rather
than 6.4~keV bolsters this interpretation. To make a more quantitative
assessment of the plausibility of our proposed explanation, we carried
out the following tests:

\begin{enumerate}

\item
We extracted the total spectrum from the largest circular region that
could reasonably fit in the \chandra\ field of view (centered on the
AGN, with a radius of 50~pixels or $24.\!\!^{\prime\prime}5$) and used
it to look for an Fe~K$\alpha$ line, as described above. In this
spectrum we detect the Fe~K$\alpha$ line at 68\% confidence, with an
equivalent width of 43~eV and we obtain an upper limit of 170~eV at
99\% confidence, which is formally consistent with the \asca\
measurement from 1998. This test illustrates that extended emission
makes an increasing contribution to the equivalent width of the line
in larger apertures.

\item
We used the spectral models for the AGN and the diffuse emission
derived from the \chandra\ observation to calculate the 2--10~keV
spectrum and the {\it apparent} equivalent width of the Fe~K$\alpha$
line. For the purposes of this calculation we assumed the flux and
spectrum of the AGN measured above and spectrum of the extended
emission determined in \S5.2 and summarized in Table~\ref{fuzztable},
adopting $Z=Z_{\odot}$ and $F_{\rm X}(2-10~{\rm keV})=2.4\times
10^{-13}~{\rm erg~s^{-1}}$, which is apropriate for the {\it entire}
circumnuclear nebula. We find that the equivalent width of the
Fe~K$\alpha$ complex in the combined AGN plus circumnuclear plasma
spectrum is approximately 40~eV. This implies that the entire strength
of this line complex measured by \asca\ can be accounted for, if we
allow the entire galaxy to have 6 times the emission measure of the
circumnuclear region, which is a reasonable possibility.

\end{enumerate}

\section{Variability of Bright Sources}

To study the variability properties of the bright discrete sources in
NGC~4736 and NGC~4579, we extracted their 0.5--8.0~keV light curves
from the same spatial regions used to extract spectra. Since only one
quarter of the S3 CCD was read out during the observations of these
two galaxies, the frame time was 0.84~s. In the case of the AGN in
NGC~4579 pileup is a cause for concern since it affects the properties
of the light curve. The qualitative effect of pileup is to dilute the
amplitude of any variability since it leads to some pairs of
lower-energy photons being counted as single, higher-energy
photons. Moreover, pileup also decreases the $S/N$ since some photon
grades are transformed to unacceptable grades and those photons are
not counted at all. The magnitude of these effects cannot be
quantified without detailed and extensive simulations, which are well
beyond the scope of this work. We have nevertheless made an effort to
minimize the effects of pileup on the light curve of the AGN by
restricting the band to 0.5--1.0~keV. This narrow, soft band
encompasses the peak of the instrument sensitivity, thus including a
significant fraction of the detected photons. The fact that it spans
only a factor of 2 in energy means that any pairs of lower-energy
photons that are counted as one will not be included in the
band. Contamination by pairs of photons below 0.5~keV counted within
this band is small because the sensitivity of the CCD declines steeply
below this energy.  Hence, the most significant effect of pileup on
light curves from this band is to reduce the $S/N$ somewhat.  Yet
another way of minimizing the effects of pileup on a light curve is to
extract the light curve after excluding the core of the PSF, where
pileup really occurs (cf, \S5.4). Thus we have also extracted a light
curve from an annulus with inner and outer radii of
$0.\!\!^{\prime\prime}5$ and $4.\!\!^{\prime\prime}2$ respectively,
which includes approximately half of the total counts.

\begin{figure}
\plotone{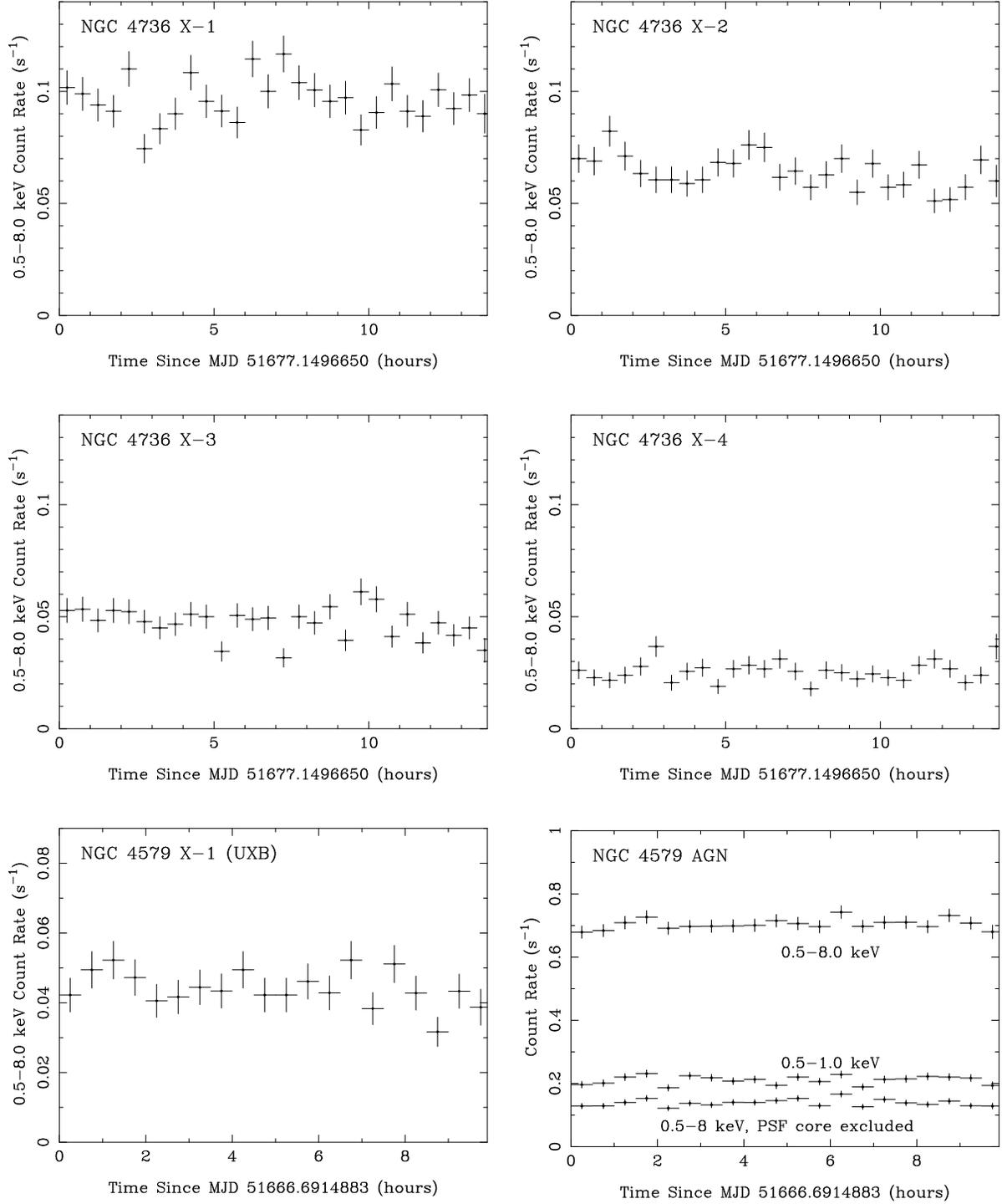}
\caption{\label{lcurves}
Light curves of bright point sources in NGC~4736 and NGC~4579. The
objects are labeled according to the conventions defined in
Tables~\ref{sourcetable4736} and \ref{sourcetable4579}. In the case of
the AGN in NGC~4579, we plot the 0.5--1.0~keV and 0.5--8.0~keV light
curves, as well as the 0.5--8.0~keV light curve extracted from the
wings of the PSF.  In all other cases we plot only the 0.5--8.0~keV
light curves.}
\end{figure}

In Figure~\ref{lcurves} we show the light curves of all the bright
discrete sources in the two galaxies, binned into 30-minute time bins.
These light curves show that the discrete sources vary on time scales
of a few hours at the 20--30\% level, while the AGN varies at the
5--10\% level.  The magnitude of the variability is quantified by
applying the excess variance test \citep{nandra97a} to these light
curves, whose results are summarized in Table~\ref{vartable}. In the
case of NGC~4579, the excess variance computed from the soft light
curve agrees with the value obtained from the outer-PSF light curve
and is a factor of 2 higher than the full-band light curve. This
suggests that pileup dilutes the variability in the full-band light
curve but not in the other two light curves. It also suggests that the
shape of the spectrum does not change as the light curve flickers. The
level of variability of the AGN in NGC~4579 is higher than what is
observed in typical LINERs of comparable luminosity \citep{p_etal98}
and approaches that of Seyfert galaxies whose luminosities are 1--2
orders of magnitude higher than that of NGC~4579. Unfortunately, it is
difficult to draw more extensive conclusions from the variability
properties of the nucleus of NGC~4579 because of the caveats
associated with pileup.

We have also searched for coherent, periodic signals in the light
curves of the bright discrete sources, but no such signals were found.
We carried out this search by computing the Fourier transforms of the
source light curves up to frequencies of 0.6~Hz, following the
methodology of \citet*{eph91}.  Using the noise properties of the
power spectra we have estimated upper limits to the amplitudes of
undetected signals, which we list in Table~\ref{vartable}. These
limits are expressed as a minimum detectable ``pulsed fraction,'' which
is the amplitude of a sinusoidal signal relative to the mean
level of the light curve.

\section{Detection and Population Properties of the Discrete X-Ray sources}

To study the population of discrete X-ray sources in the target
galaxies, we used the source detection program \verb+WAVDETECT+
\citep{freeman01}, which is included in the \verb+CIAO+ package.  We
searched for sources on several spatial scales (from 1.4 to 11.6
pixels, in steps of $\sqrt{2}$), accepting detections associated with
less than $10^{-6}$ chance probabilities due to local background
fluctuations.  We repeated the search twice, first in the full,
0.5--8~keV, band and then in the soft, 0.5-2~keV, band. We combined
the results of the two passes to compute an X-ray color for each
source, which we define as the ratio of counts in the hard (2--8~keV)
band to counts in the soft (0.5--2~keV) band. Because the sensitivity
of the instrument rises steeply towards low energies, all of the
sources are detected in the total and soft bands, but nearly half of
the sources have a hard band count rate consistent with zero. In the
end our detection limit corresponds to about 6 counts per source,
although it varies over the field of view because of (a) vignetting,
(b) degradation of the image quality off axis, and (c) patchy diffuse
emission, which increases the effective local background. After
carrying out the above automatic procedure, we checked the measured
properties of many of the detected sources manually. We paid
particular attention to sources in the crowded field in the nucleus of
NGC~4736 and we excluded any sources falling along the readout streak
in the image NGC~4579.  We tabulate the properties of detected sources in
Tables~\ref{sourcetable4736} and \ref{sourcetable4579}, where we give
the coordinates of each source, the total count rate, the X-ray color,
and an estimate of the 0.5--8~keV luminosity. For sources whose hard
count rate is consistent with zero, we only give an upper limit to
the color.

To estimate the luminosities given in the source tables we assumed
that the source spectra could be described by a simple power law with
an absorbing column equal to the Galactic column. With this assumption
we simulated spectra with a wide range of photon indices (0.5--5.0)
and measured and tabulated their colors and count rates in the same
way as we did for the observed spectra. We used the resulting lookup
table to estimate the photon indices and fluxes of detected sources
from their observed colors and count rates. By comparing the results
of this method with results obtained for bright sources for which
spectra were extracted and fitted with models, we estimate the
uncertainty in the photon index and flux to be about 10--20\%.  This
uncertainty is in addition to the Poisson uncertainty on the count rate
listed in Tables~\ref{sourcetable4736} and \ref{sourcetable4579}.  In
cases when only an upper limit to the color was determined from the
data, we used this upper limit to determine the photon index. Thus the
uncertainty in the flux of these sources can be up to 50\%, judging
from the relation between photon indices and count rates in simulated
spectra.

As we already mentioned in \S4, in NGC~404 we do not find any X-ray
sources other than the nucleus, down to a limiting luminosity of
$1\times 10^{36}~{\rm erg~s^{-1}}$. In NGC~4736 and NGC~4579 we find
respectively 39 and 21 sources, with luminosities in the range
$1\times 10^{36} - 1\times 10^{39}~{\rm erg~s}^{-1}$ in the former
galaxy and $1\times 10^{37} - 9\times 10^{39}~{\rm erg~s}^{-1}$ in the
latter. The lower end of the range is effectively our detection
limit. Thus, in the case of NGC~4736 we can detect typical supernova
remnants (SNRs) and Be-star X-ray binaries (BeXBs), while in the case
of NGC~4579, we can detect low-mass X-ray binaries (LMXBs) or more
luminous objects. We have estimated the probability of contamination
of the source population by unrelated, background sources using the
number counts of sources derived from deep \chandra\ surveys by
\citet{t_etal01}. In the observation of NGC~4736 the limiting flux is
$4.8\times 10^{-16}~{\rm erg~cm^{-2}~s^{-1}}$, which yields an
expectation value of 5 background sources within the field of view,
while in the observation of NGC~4579 the limiting flux is $5.6\times
10^{-16}~{\rm erg~cm^{-2}~s^{-1}}$ and the expected background source
number is 4. These are conservative upper limits to the number of
background sources, since the flux of such sources would be attenuated
as it passes through the disk of the foreground galaxy. If we assume,
for example, that the column density perpendicular to the galactic
disks is $1\times 10^{21}~{\rm cm}^{-2}$, the expected number of
background sources in each case decreases by 1. Thus, contamination
from background sources is negligible from the point of view of our
statistical studies, although it may be an issue in our discussion of
the properties of individual sources. We also considered the
possibility of foreground contamination, i.e. that some of the
detected X-ray sources are foreground stars. The absence of optical
counterparts of any of the detected X-ray sources (see \S4) indicates
that there is no foreground contamination. This is not unexpected
since the Galactic lattitude of the target galaxies is rather high
($-27^{\circ}$, $+76^{\circ}$, and $+74^{\circ}$ for NGC~404,
NGC~4736, and NGC~4579 respectively). The faintest optical objects
that could be associated with detected X-ray sources are M
stars. Using the X-ray-to-optical flux ratios of M stars reported by
\citet{h_etal95}, we estimate that M star counterparts to the sources
that we have detected should have $R<20.2$, and would have been
included in the USNO~A2.0 catalog.

The spatial distribution of sources differs markedly between NGC~4736
and NGC~4579. In the former galaxy there is a concentration of sources
within 10\arcsec\ from the nucleus, and a separate concentration at
the radius of the star-forming ring ($\sim 45$--50\arcsec).  Thus, the
distribution of X-ray sources does not follow the distribution of
infrared light (i.e., starlight) from the host galaxy, but rather the
distribution of emission-line light \citep[cf,][]{s_etal94},
suggesting a connection with star formation. In contrast, the sources
in NGC~4579 are distributed smoothly with distance from the center.

\begin{figure}
\plotone{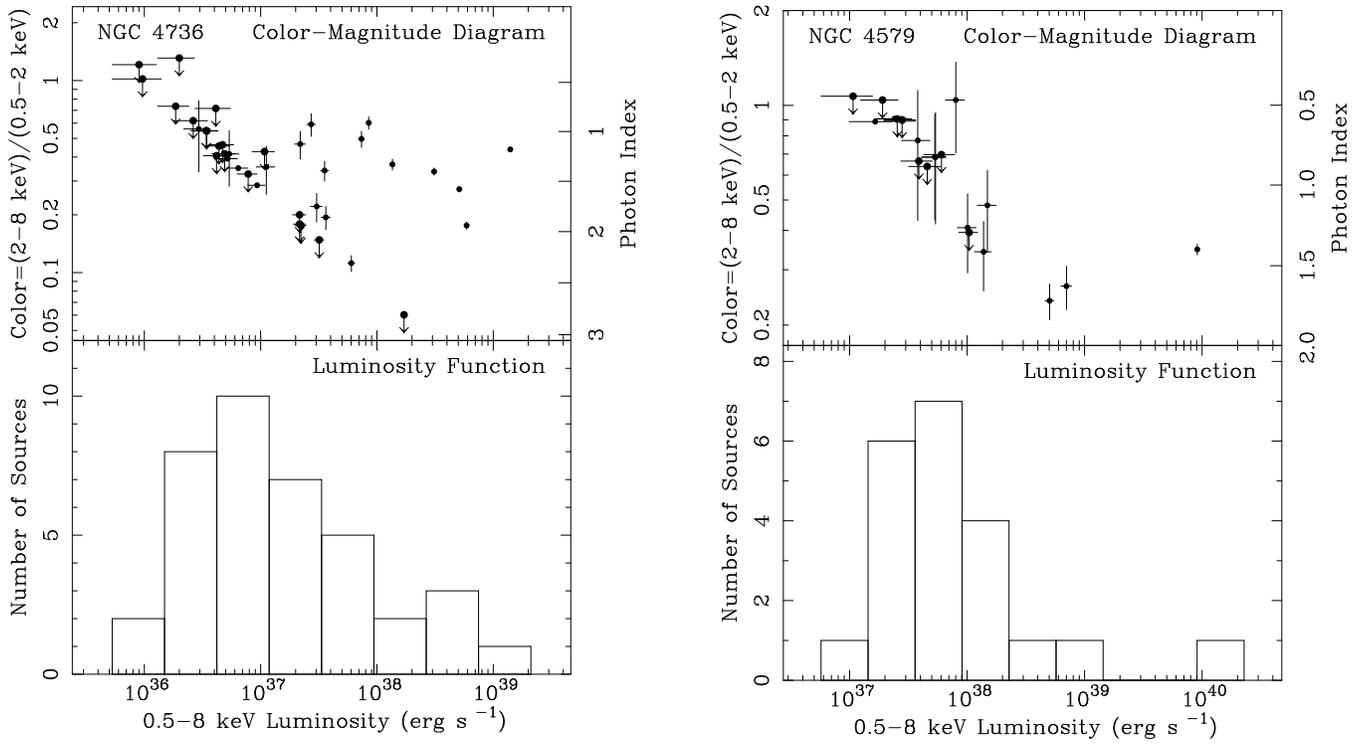}
\caption{\label{colormag} Color-Magnitude diagrams and luminosity
distributions for the X-Ray binaries in NGC~4579 and NGC~4736. In the top
frame of each set the vertical axis is calibrated in color (left) and
in spectral index (right). Upper limits to the color are denoted by
arrows.}
\end{figure}

In Figure~\ref{colormag} we show the ``color-magnitude'' diagrams and
the luminosity distributions of the source populations of the two
galaxies. In both galaxies, the most luminous sources have photon
indices between 1.0 and 2.0. At lower luminosities only limits on the
photon index are available, although there are some notable exceptions
in both galaxies, where we detect rather hard sources ($\Gamma\sim 1$)
at $L_{\rm X}\sim 10^{37}~{\rm erg~s^{-1}}$. The luminosity
distribution appears to peak just below $10^{37}~{\rm erg~s^{-1}}$ in
NGC~4736 and just below $10^{38}~{\rm erg~s^{-1}}$ in NGC~4579.

In Figure~\ref{lumfunc} we compare the cumulative X-ray source
luminosity functions of NGC~4736 and NGC~4579 with those of other
galaxies with published \chandra\ data. We include two early-type
galaxies, NGC~4697 and NGC~1553 \citep[E and S0,
respectively;][]{sib01,bsi01} and two starbursting spirals, M82 and
the Circinus galaxy \citep{g_etal00,bauer_etal01}.  NGC~4579,
NGC~4697, and NGC~1553 stand out in this figure because their
cumulative X-ray source luminosity functions are steeper than those of
the other three galaxies.  This difference can be quantified by
fitting a simple power-law model [$N(L)\propto L^{-\alpha}$ or
$N(>L)\propto L^{-(\alpha-1)}$] to the data at luminosities above the
initial plateau. In the case of NGC~4579, NGC~4697, and NGC~1553 we
find power-law indices ($\alpha$) between 2.0 and 2.4, while in the
other two galaxies the power-law indices range between 1.4 and 1.6. In
fact, the difference is more extreme than what the above power-law
indices indicate because the X-ray luminosity functions of the two
early-type galaxies steepen at luminosities higher than about
$10^{38}~{\rm erg~s}^{-1}$ \citep{sib01,bsi01}.

It is interesting to note that the galaxies with steep X-ray source
luminosity functions are the ones without signs of current or recent
star formation activity. In this context it is noteworthy that a
similar effect is observed in two nearby spiral galaxies, M31 and M81
\citep*{pfj93,tennant01}, where the X-ray luminosity functions of
bulge and disk sources have been measured separately. The bulge
luminosity functions resemble those of early type galaxies, while the
luminosity function of the disk of M81 resembles that of the
starbursting spirals (the disk luminosity function of M31 has not been
reported).  In \S{8} we interpret this difference in X-ray source
luminosity functions by appealing to current or recent intense star
formation and the age of the underlying stellar population.

\section{Discussion}

\subsection{Implications for the Central Engines of LINERs}

Our original aim in undertaking this program was to search for the
power source of the optical emission lines, which are the hallmark of
LINERs.  Our main result is that the objects in our small collection
are quite diverse in their properties, with a different power source
apparently playing the dominant role in each galaxy. As a first step
in considering the implications of these findings, we compare them
with published results of previous X-ray observations of these
galaxies with \asca. In Table~\ref{budgettable} we list the
0.5--10~keV luminosity of the different components contributing to the
X-ray emission from each galaxy. This table quantifies the general
description given in \S4. The sum of these components reproduces
fairly closely the flux detected by \asca\ from each of these
galaxies. One notable discrepancy between the \chandra\ and \asca\
spectra is the detection of Fe~K$\alpha$ lines at 6.4 and 6.7~keV in
the \asca\ spectra of both NGC~4736 and NGC~4579
\citep{rwo99,t_etal98}. Such lines are not seen in the \chandra\
spectra of any of the bright discrete sources in these galaxies. What
\chandra\ does detect, however, is a hot component
($kT\sim\;$~several~keV) to the plasma that permeates the central
regions of these two galaxies. The discrepancy can thus be resolved if
the Fe~K$\alpha$ emission lines in the \asca\ spectra arise from this
hot plasma, as discussed in \S5.4.

In NGC~4579 the case for an AGN seems iron clad. A central, compact
source dominates the power budget, as we illustrate in
Table~\ref{budgettable}. The spectral energy distribution of this
source has a significant contribution from UV light, broad optical and
UV emission lines, as well as a non-thermal radio spectrum with a
brightness temperature of $2\times 10^8$~K \citep{f_etal00}. All these
properties are classic traits of an AGN.

At the opposite extreme, NGC~404 harbors a rather anemic star-forming
region in its nucleus. Its X-ray luminosity of $1\times 10^{37}~{\rm
erg~s^{-1}}$ is comparable to that of a single HMXB or a giant
star-forming region such as 30 Doradus \citep[cf,][]{wh91}.  The soft,
thermal X-ray spectrum favors an origin of the X-rays in the hot gas
of a superbubble, powered primarily by off-center supernovae
\citep[see the discussion by][]{cm90}. This view is supported by the
absorption-line UV spectrum and the monochromatic luminosity at
2270~\AA\ of $L_{\lambda} (2270~{\rm \AA}) = 1.2\times 10^{36}~{\rm
erg~s^{-1}~\AA^{-1}}$ \citep[from][]{m_etal95}, which is equivalent to
that of 10 late-O supergiants. The sound speed in the hot gas is of
order $c_{\rm s} \approx 380\; (kT/{\rm 0.5\; keV})^{1/2}~{\rm
km~s^{-1}}$, which is considerably larger than the velocity dispersion
of the stars in the nucleus of the galaxy \citep[$\sigma_*=55~{\rm
km~s^{-1}}$;][]{f_etal89}. Thus the gas must be unbound, a conclusion
bolstered by the ``blow-out'' morphology of the H$\alpha$+[\ion{N}{2}]
image, which also lends support to the superbubble hypothesis for the
X-ray emitting gas.

In the case of NGC~4736, there is no compelling evidence for the
presence of an AGN even though this galaxy is fairly bright in X-rays
and harbors an unresolved UV source. The X-rays are produced by a
dense cluster of stellar X-ray sources in the inner 2~kpc of the
galaxy, and to a lesser degree by diffuse emission from hot gas (see
the energy budget in Table~\ref{budgettable}). This possibility was
correctly anticipated and discussed by \citet{rwo99} and \citet{rsw01}
who nonetheless preferred an AGN interpretation in the end. As we
argue below, these X-ray sources could have been produced in a recent
episode of star formation and probably consist of high-mass X-ray
binaries (HMXBs), LMXBs,and SNRs. The brightest nuclear X-ray sources
have very similar X-ray spectra and comparable luminosities, making
the association of any one of them with an AGN rather implausible.  It
should be pointed out, however, that source X-2, coincides with the
nuclear UV source detected by the {\it HST}; thus it stands out among
the nuclear X-ray sources as the only one with a UV counterpart.  But
even if we associate this source with an AGN, it is not very important
energetically since it contributes less than 20\% of the X-ray
luminosity and less than half of the UV luminosity.\footnote{For the
sake of completeness, we note that the stellar velocity dispersion of
NGC~4736 of 136~km~s$^{-1}$ \citep{mcelroy95} implies a black hole
mass of $2-3\times 10^7~{\rm M}_{\odot}$ and an Eddington ratio of
$L/L_{\rm Edd}\approx 3\times 10^{-6}$ (see \S8.4 for the details of
the methodology).} The properties of the non-thermal radio continuum
emission from the nucleus \citep{th94} are consistent with an origin
in a group of several, young SNRs of the Cas~A variety, whose kinetic
energy heats the gas responsible for the diffuse X-ray emission \citep
[cf, the discussion of the diffuse X-ray emission of the LMC by][]
{whhw91}.  In the same spirit, the nuclear, point-like, UV source is
very likely an OB association, which hosts a HMXB (NGC~4736 X-2).  The
observations of \citet{m_etal95} provide a lower limit to the
monochromatic luminosity at 2270~\AA\ (the central point source is
saturated) of $L_{\lambda} (2270~{\rm \AA}) > 4.4\times 10^{36}~{\rm
erg~s^{-1}~\AA^{-1}}$, which is equivalent to the luminosities of 5
late-O supergiants. Independent evidence from optical spectroscopy by
\citet{tani96} leads to similar conclusions about the nature of the
activity in the nulceus of NGC~4736. These authors find that the
optical absorption-line spectrum of the nucleus of NGC~4736 agrees
with models of a $10^9$-year old starburst. They also find that the nuclear
H$\alpha$ luminosity is no larger than that of luminous Galactic
\ion{H}{2} regions and that six O6 stars are enough to power it. Thus,
this particular LINER seems to be associated with a recent or aging
starburst rather than an AGN.

\subsection{The Discrete Source Populations} 

The discrete sources detected in NGC~4736 and NGC~4579 may include a
variety of objects. The most luminous sources ($L_{\rm X}>10^{37}~{\rm
erg~s}^{-1}$) are most likely X-ray binaries: a combination of LMXBs
and HMXBs. This conclusion is supported by two pieces of evidence:
their spectra, which are similar to those of bright Galactic
X-ray binaries, and their variability properties, which suggest that
these are individual objects, rather than clusters of less luminous
sources.  Other types of object, such as SNRs and BeXBs, could
contribute to the population of lower luminosity sources, detected
primarily in NGC~4736. If we could determine the colors of the
lower-luminosity sources we could perhaps use them to differentiate
between SNRs and BeXBs since the 2--10~keV spectra of the former are
dominated by line emission at low energies, while the spectra of the
latter resemble power laws.

\begin{figure}
\plotone{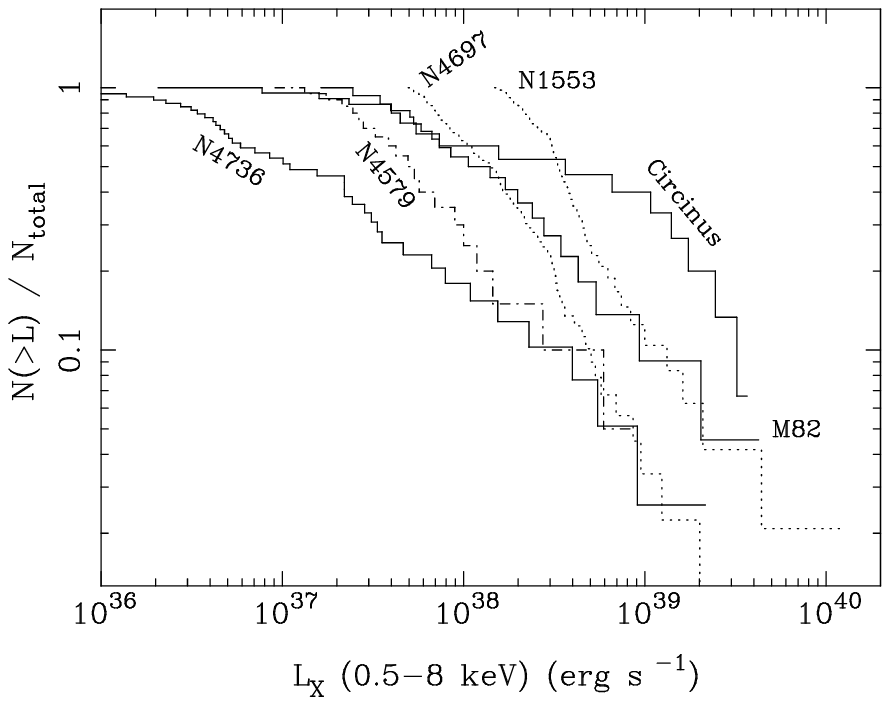}
\caption{\label{lumfunc} Cumulative X-ray source luminosity functions of 
NGC~4579 (dot-dashed line) and NGC~4736 (solid line) compared to those
of other galaxies observed by \chandra. The comparison galaxies are
NGC~4697 and NGC~1553 (E and S0, dotted lines), and M82 and Circinus
(starbursting spirals, solid lines). The UXB in NGC~4579 is excluded
from its luminosity function in the interest of clarity.}
\end{figure}

The high concentration of luminous sources in the nucleus of NGC~4736
suggests that we are observing the aftermath of an intense episode of
star formation.  By extension, a significant fraction of the discrete
sources in NGC~4736 could be HMXBs rather than LMXBs. Models of
populations of interacting binaries associated with bursts of star
formation \citep*{cmk97,vb00,se00} show an initial hard-X-ray luminous
phase, which is intimately connected with the rise and fall of HMXBs
\citep[see also the discussion of the processes by][]{dalton95}.  The
first HMXBs make their appearance after the death of the most massive
O stars, and shine brightly until the demise of their O and B
secondaries. At this point the total hard X-ray luminosity of the
population should decline until the first luminous LMXBs arrive on the
scene, causing the X-ray luminosity to level off with time. Thus, the
hard X-ray emission from the population should persist for up to
50~Myr after the vigorous star formation has stopped. Inspired by the
model predictions, we propose that the nucleus of NGC~4736 is an
``aging'' starburst or in an immediate post-starburst phase, where the
HMXBs dominate the top of the X-ray source luminosity function.  In
support of our hypothesis we note the qualitative difference between
the cumulative luminosity functions of X-ray sources in nearby
galaxies, shown in Figure~\ref{lumfunc} and discussed in \S7. Galaxies
currently undergoing vigorous star formation, such as M82 and
Circinus, have cumulative luminosity functions that are flatter than
those of galaxies with primarily old stellar populations such as
NGC~4697 and NGC 1553. We speculate that the flatter slopes result
from an excess of luminous HMXBs that have formed recently. The steep
luminosity function of NGC~4579 fits well within this picture since
there are no indications of ongoing, intensive star formation in its
bulge or disk.

A particularly striking source is NGC~4579 X-1, which could be a
UXB. If this object is at the distance of NGC~4579, as we have been
implicitly assuming, its 2--10~keV luminosity would exceed
$10^{40}~{\rm erg~s}^{-1}$. Its spectrum is very similar to the
spectra of the X-ray binaries in NGC~4736 and, in fact, it is the only
one of the bright X-ray sources whose spectrum is better described by
a Comptonized blackbody model than by a simple power law.  This object
is consequently very similar to luminous Galactic X-ray binaries. Its
luminosity corresponds to the Eddington limit for a 70~M$_{\odot}$
object, which could be regarded as a lower limit on the mass of the
accreting compact object, at first glance. However, the Eddington
limit need not apply in the non-spherical accretion geometry of X-ray
binaries, and a number of cases are known in which it is blatantly
violated. There are at least 3 X-ray pulsars in the Magellanic Clouds
\citep*[LMC~X-4, SMC~X-1, and A~0538--66;][and references
therein]{nagase89,woo95} whose luminosities exceed the Eddington limit
for a neutron star by as much as a factor of 5 \citep[and even by a
factor of 30 in the flares of LMC~X-4;][]{woo95}. Of course violations
of the Eddington limit may be transient \citep[cf, the discussion of
SMC~X-1 by][]{hm01} or merely appear as such because of beaming
\citep{king01}. However, the fact remains that at certain times or
from certain viewing directions some X-ray binaries appear extremely
luminous. Thus, the accreting compact object in NGC~4579~X-1 need not
be more massive than about 10~M$_{\odot}$. Another possibility is that
this source is at a large distance behind NGC~4579. The fact that we
expect 3 background sources in the image of NGC~4579, as well as the
large measured column density (comparable to the column density
perpendicular to the disk of a spiral galaxy) make this a plausible
scenario. It should be pointed out, however, that the 3 expected
background sources should be rather faint according to the flux
distributions measured in deep surveys.  The {\it a priori}
probability of finding a background object with the 2--10~keV flux of
NGC~4579 X-1 in our field of view is approximately 6\%. If this source
were a background AGN, then its redshift, as inferred from the
observed flux, would be
$ z=0.01\; \left({H_0/50~{\rm km~s^{-1}~Mpc^{-1}}}\right)\;
\left({L_{\rm X} / 10^{42}~{\rm erg~s^{-1}}}\right)^{1/2}, $
suggesting that the source could easily be a garden variety Seyfert
galaxy at $z=0.01$ or a luminous quasar at $z=1$. Ascertaining the
nature of this source will have to await identification of its optical
counterpart.

\subsection{The AGN and Extended Emission in NGC~4579}

Using the uncontaminated luminosity of the AGN in NGC~4579, we can
explore the properties of the accretion flow onto the supermassive
black hole that powers it. We can estimate the mass of the central
black hole using recently established correlations between it and the
velocity dispersion of stars in its immediate vicinity
\citep{gebhardt00,mf01}. The velocity dispersion of $185~{\rm
km~s}^{-1}$ measured by \citet{dressler87} yields a black hole mass in
the range 6 to $13\times 10^7~{\rm M}_{\odot}$, comparable to what is
found in Seyfert galaxies \citep{f_etal01}. Assuming that the observed
0.5--10~keV luminosity is about 10\% of the bolometric luminosity, we
can estimate the Eddington ratio for this AGN to be $L/L_{\rm
Edd}\approx 1-2\times 10^{-4}$, which is a low value, consistent with
expectations for a low-luminosity AGN.  At such a low Eddington ratio,
the inner accretion flow is expected to be advection-dominated (an
ADAF or similar structure) and thus very hot, vertically extended, and
optically thin. Such a structure would be very different from a
conventional, optically thick, geometrically thin accretion disk in
its observational signature. In particular, the absence of an
optically thick inner accretion disk can explain the weakness of the
Fe~K$\alpha$ line of NGC~4579 relative to what is observed in Seyfert
galaxies \citep*[see the detailed discussion in][]{esm00}. A
vertically extended ADAF comprising the inner accretion structure can
illuminate the outer thin disk and drive the emission of optical
Balmer lines as well as X-ray Fe~K$\alpha$ lines. This scenario offers
an appealing explanation for the double-shouldered profile of the
H$\alpha$ line of NGC~4579 \citep{barth_etal01} just as it does for
the double-peaked Balmer lines in broad-line radio galaxies
\citep{eh94}. However, models for the continuum spectra of ADAFs
\citep*[e.g.,][]{nym96} predict X-ray spectral indices flatter than
the typical values observed in Seyfert galaxies.  This does not seem
to hold for NGC~4579, where the 2--10~keV spectral index is typical of
a Seyfert galaxy but the equivalent width of the Fe~K$\alpha$ line is
not. However, the UV-X-ray spectral index ($\alpha_{\rm ox}$) of
NGC~4579 is rather typical of LINERs as shown by \citet{h99}.
Interestingly enough, the variability amplitude is also high for a
low-luminosity AGN and is comparable to what is observed in Seyferts
(see \S6). Therefore, NGC~4579 appears to be a ``hybrid'' object with
some properties characterisitc of Seyferts and other characteristic
of LINERs or low-luminosity AGNs. 

The heating mechanism of the extended emission region of NGC~4579 is
unclear.  We can confidently exclude any mechanical interaction
between it and an AGN jet since the VLBI radio images of
\citet{f_etal00} show only an unresolved radio core with no extension.
One possible mechanism is that the extended emission region is
photoionized by the AGN. The X-ray luminosity of the AGN is 20 times
higher than that of the circumnuclear nebula and a fraction of the
hard X-ray luminosity from the circumnuclear region may be contributed
by BeXBs and LMXBs, which we have not detected.  Thus, photoinization
may be a viable mechanism.  Moreover, as \citet{m_etal95} point out,
the extrapolation of the observed UV luminosity is enough to power the
optical emission lines from the same region.  Another mechanism is
heating by stellar processes in the wake of a circumnuclear
starburst. In the context of this scenario, the starburst should be
old enough that its luminous O stars have passed away and supernova
remnants have blended together into a diffuse circumnuclear medium.
Unfortunately, we cannot assess the likelihood of the above scenarios
with the information currently available.

\section{Epilogue}

The results of our exploratory observations indicate that \liners\ are
a heterogeneous population of objects. This should not come as a
surprise, in view of the way that the class is defined. But an
important question that remains outstanding is what fraction of
\liners\ are powered by accretion onto a supermassive black hole and
how these objects can be identified unambiguously. Unfortunately, the
first part of this question cannot be answered until the second part
is answered first. One rather clear signature of an AGN is the
presence of broad emission lines in the optical or UV spectra of
\liners. However, the absence of broad emission lines cannot be
taken as evidence that there is no AGN. This is especially true when
one considers the practical difficulties of finding weak broad lines
in the starlight-dominated spectra of \liners.  Another possibly clear
signature of an AGN is the presence of a non-thermal radio core with a
high brightness temperature.

X-ray observations can also shed light on this question, although they
are more demanding in terms of resources and may not always yield a
definitive answer. The high spatial resolution of \chandra\ now allows
us to examine the morphology of \liner\ nuclei in detail and combine
morphological and spectral information, just as we have done
here. However, one can also envisage cases where a \liner\ harbors a
single, discrete source of ``intermediate'' luminosity
($10^{39}$--$10^{40}~{\rm erg~s^{-1}}$), which could equally well be
an AGN or a UXB. Moreover, observations with {\it ASCA} should be
regarded with caution, since they can sometimes be misleading because
of their poor spatial resolution. A good case in point is provided by
NGC~4736, whose X-ray binaries and diffuse emission produce a
spatially integrated spectrum that can be modeled by a hard power law
and a thermal plasma component. This ``canonical'' \liner\ spectrum
can result from a current or recent starburst, just as well as it
can result from an AGN.

\acknowledgements This work was supported by NASA through grant
GO0-1152A,B from the Smithsonian Astrophysical Observatory. E.C.M.  is
supported by NASA through \chandra\ Fellowship grant PF-10004, awarded
by the \chandra\ X-ray Center, which is operated by the Smithsonian
Astrophysical Observatory for NASA under contract NAS8-29073. We are
grateful to Ann Hornschemeier for many helpful discussions and
technical advice. We thank Lars Bildsten, Bob Rutledge, and Dani Maoz
for useful suggestions.  We also owe thanks to Rick Pogge and Dani
Maoz for providing us with optical images of our target galaxies in a
convenient, electronic format, and to Franz Bauer for doing the same
with the luminosity functions of X-ray siurces sources in nearby
galaxies.


\clearpage

\begin{deluxetable}{lcccccc}
\tablenum{1}
\tablewidth{6in}
\tablecolumns{6}
\tablecaption{Targets and Their Basic Properties\label{targtable}}
\tablehead{
\colhead{} &
\colhead{} &
\colhead{} &
\colhead{Spatial} &
\colhead{Galactic} &
\colhead{Galactic} &
\colhead{}
\cr
\colhead{Target} &
\colhead{Hubble} &
\colhead{Distance\tablenotemark{\; b}} &
\colhead{Scale} &
\colhead{$N_{\rm H}$\tablenotemark{\; c}} &
\colhead{$E(B-V)$\tablenotemark{\; d}} &
\colhead{$v_{\odot}$\tablenotemark{\; e}} 
\cr
\colhead{Name(s)} &
\colhead{ Type\tablenotemark{\; a}} &
\colhead{(Mpc)} &
\colhead{(pc/$^{\prime\prime}$)} &
\colhead{($10^{20}~{\rm cm}^{-2}$)} &
\colhead{(mag)} &
\colhead{(km s$^{-1}$)} 
}
\startdata
NGC~404        & SA(s)0--:  &  2.4 & 12 & 5.30 & 0.059 & --46  \cr
NGC~4736 (M94) & (R)SA(r)ab &  4.3 & 21 & 1.41 & 0.018 &  307  \cr
NGC~4579 (M58) & SAB(rs)b   & 16.8 & 81 & 2.54 & 0.041 & 1521  
\tablerefs{
(a) \citet{RC3},
(b) \citet{tully88},
(c) \citet{stark92},
(d) \citet*{sfd98},
(e) \cite{hfs_a97}}
\enddata
\end{deluxetable}


\begin{deluxetable}{lll}
\tablenum{2}
\tablewidth{3.5in}
\tablecolumns{3}
\tablecaption{Observation Log\label{obstable}}
\tablehead{
\colhead{} &
\colhead{Observation} &
\colhead{Exposure} 
\cr
\colhead{} &
\colhead{Start} &
\colhead{Time} 
\cr
\colhead{Target} &
\colhead{(UT)} &
\colhead{(s)} 
}
\startdata
NGC~404  & 1999 Dec 19, 21:12:43 & 23,864 \cr
NGC~4736 & 2000 May 13, 03:12:11 & 47,366 \cr
NGC~4579 & 2000 May 02, 16:26:44 & 33,905 
\enddata
\end{deluxetable}


\begin{deluxetable}{lcccc}
\tablenum{3}
\tablewidth{6.5in}
\tablecolumns{4}
\tablecaption{Parameters of Thermal Plasma Models\label{fuzztable}}
\tablehead{
\colhead{} &
\colhead{} &
\colhead{} &
\multicolumn{2}{c}{NGC 4579\tablenotemark{\; b}} \cr
\colhead{} &
\colhead{} &
\colhead{} &
\multicolumn{2}{c}{\hrulefill} \cr
\colhead{Model Parameters} &
\colhead{NGC 404\tablenotemark{\; a}} &
\colhead{NGC 4736\tablenotemark{\; b}} &
\colhead{$Z=$free} &
\colhead{$Z=Z_{\odot}$} 
}
\startdata
$N_{\rm H}~(10^{20}~{\rm cm}^{-2})$ & $<6$                & $1.4^{\; c}$         & $<6$                & $7^{+9}_{-4}$ \cr
Abundance ($Z/Z_{\odot}$)           & $<0.12$             & $0.3^{+1.4}_{-0.2}$ & $0.2^{+1.3}_{-0.1}$ & 1 (fixed) \cr
$kT^{\; d}$ (keV)                   & $0.6^{+0.3}_{-0.4}$ & $0.56\pm 0.06$      & $0.3\pm 0.2$, $0.6^{+0.4}_{-0.1}$, $>6$ & $0.11^{+0.05}_{-0.04}$, $0.58^{+0.04}_{-0.08}$, $10^{+44}_{-4}$ \cr
$EM^{\; e}~(10^{61}~{\rm cm}^{-3}$) & 0.35                & 0.7                 & 33, 34, 28          & 34, 11, 27 \cr
Total $\chi^2$                      & 10.45               & 3.159               & 22.507              & 25.546 \cr
Reduced $\chi^2$                    & 0.696               & 0.632               & 1.072               & 1.161 \cr
\sidehead{Flux $({\rm erg~cm^{-2}~s^{-1}})$}
0.5-2~keV                           & $1.5 \times 10^{-14}$ & $2.0\times 10^{-14}$ & $1.2\times 10^{-13}$ & \cr
2-10~keV                            & \dots\                &  ...                 & $1.2\times 10^{-13}$ & \cr
\sidehead{Luminosity $({\rm erg~s^{-1}})$}
0.5-2~keV                           & $1.0 \times 10^{37}$  & $4.4\times 10^{37}$  & $4.1\times 10^{39}$  & \cr
2-10~keV                            & \dots                 &  ...                 & $4.1\times 10^{39}$  & \cr
\enddata
\tablenotetext{a\;}{The spectrum was extracted from a circular region of area 13~square arcseconds, equal to
the area of a point source.}
\tablenotetext{b\;}{The spectrum was extracted from a region of area 239~square arcseconds.}
\tablenotetext{c\;}{Held fixed at the Galactic value because it was unconstrained.}
\tablenotetext{d\;}{Single-temperature models for NGC~404 and NGC~4736, 3-temperature model for NGC~4579.}
\tablenotetext{e\;}{Emission measure: $EM = \int n_{\rm e} n_{\rm i}\; dV$, where $n_{\rm e}$ and $n_{\rm i}$ are the 
electron and ion densities.}
\end{deluxetable}


\begin{deluxetable}{cclcccl}
\tablenum{4}
\tablewidth{5.5in}
\tablecolumns{7}
\tablecaption{Power-Law Fits to Spectra of Bright Discrete Sources\label{powertable}}
\tablehead{
\colhead{} &
\colhead{Column} &
\colhead{Power-Law} &
\colhead{Observed} &
\colhead{2--10 keV} &
\colhead{} &
\colhead{} \cr
\colhead{} &
\colhead{Density} &
\colhead{Photon} &
\colhead{2--10 keV Flux} &
\colhead{Luminosity\tablenotemark{\ a}} &
\colhead{Total} &
\colhead{Reduced} \cr
\colhead{Object} &
\colhead{($10^{20}~{\rm cm}^{-2}$) } &
\colhead{Index} &
\colhead{(${\rm erg~cm^{-2}~s^{-1}}$)} &
\colhead{(${\rm erg~s^{-1}}$) } &
\colhead{$\chi^2$} &
\colhead{$\chi^2$}
}
\startdata
\sidehead{\it NGC~4736}
X-1 & $<3.7$ & $1.13\pm 0.10$ & $6.4\times 10^{-13}$ & $1.4\times 10^{39}$ & 34.13 & 0.7937 \cr
X-2 & $<3.3$ & $1.6\pm 0.1$ & $2.7\times 10^{-13}$ & $5.9\times 10^{38}$ & 20.92 & 0.9511 \cr
X-3 & $10^{+13}_{-9}$ & $1.8^{+0.7}_{-0.4}$ & $2.3\times 10^{-13}$ & $5.1\times 10^{38}$ & 30.19 & 1.078 \cr
X-4 & $<12$ & $1.6\pm 0.2$ & $1.4\times 10^{-13}$ & $3.1\times 10^{38}$ & 23.08 & 0.9234 \cr
\sidehead{\it NGC~4579}
X-1 & $18^{+6}_{-4}$ & $1.8\pm 0.2$ & $2.7\times 10^{-13}$ & $9.1\times 10^{39}$ & 22.05 & 0.9588 \cr
\tablenotetext{a\;}{Intrinsic luminosity, corrected for absorption.}
\enddata
\end{deluxetable}


\begin{deluxetable}{lcc}
\tablenum{5}
\tablewidth{5.in}
\tablecolumns{3}
\tablecaption{Variability Properties of Bright X-Ray Sources\label{vartable}}
\tablehead{
\colhead{} &
\colhead{Excess Variance} &
\colhead{Maximum Pulsed Fraction} \cr
\colhead{Object} &
\colhead{(s$^{-1}$)} &
\colhead{(\%)}
}
\startdata
\sidehead{\it NGC~4736}
X-1 & $0.06\pm 0.02$   & 13 \cr
X-2 & $0.06\pm 0.04$   & 10 \cr
X-3 & $0.10\pm 0.07$   & 17 \cr
X-4 & $<0.1$           & 19 \cr
\sidehead{\it NGC~4579}
AGN (0.5--1 keV) & $0.035\pm 0.008$ & ... \cr
AGN (0.5--8 keV) & $0.017\pm 0.001$ & ... \cr
AGN (0.5--8 keV), PSF wings & $0.05\pm 0.02$ & ... \cr
X-1 (UXB) & $<0.02$          & 18 \cr
\enddata
\end{deluxetable}


\begin{deluxetable}{lccccc}
\small
\tablenum{6}
\tablewidth{6.3in}
\tablecolumns{6}
\tablecaption{Catalog of Discrete Sources in NGC 4736\label{sourcetable4736}}
\tablehead{
\colhead{Official Name} &
\colhead{} &
\colhead{Coordinates} &
\colhead{Count Rate} &
\colhead{} &
\colhead{$L_{\rm X}$(0.5--8~keV)} 
\cr
\colhead{(CXOU J)} &
\colhead {Alias} &
\colhead{(J2000)} &
\colhead{($10^{-2}~{\rm s}^{-1}$)} &
\colhead{Color} &
\colhead{($10^{37}~{\rm erg~s}^{-1}$)} 
}
\startdata
$125041.7+410556$ &     & 12:50:41.8~~$+41$:05:56 & $0.035 \pm 0.009$ & $0.6 \pm 0.2$     &   0.29 \cr  
$125047.6+410512$ &     & 12:50:47.6~~$+41$:05:12 & $0.35 \pm 0.03$   & $0.34 \pm 0.04$   &   3.50 \cr  
$125047.7+410707$ &     & 12:50:47.7~~$+41$:07:08 & $0.05 \pm 0.01$   & $ < 0.4$          &   0.52 \cr  
$125048.6+410742$ &     & 12:50:48.6~~$+41$:07:42 & $0.51 \pm 0.03$   & $0.11 \pm 0.01$   &   6.00 \cr  
$125048.8+410635$ &     & 12:50:48.8~~$+41$:06:36 & $0.010 \pm 0.006$ & $ < 1.2$          &   0.07 \cr  
$125049.9+410640$ &     & 12:50:50.0~~$+41$:06:41 & $0.09 \pm 0.02$   & $ < 0.3$          &   0.93 \cr  
$125050.2+410752$ &     & 12:50:50.3~~$+41$:07:53 & $0.06 \pm 0.01$   & $ < 0.4$          &   0.64 \cr  
$125050.3+410712$ & X-4 & 12:50:50.3~~$+41$:07:12 & $2.67 \pm 0.08$   & $0.34 \pm 0.01$   &  31.00 \cr  
$125050.3+410628$ &     & 12:50:50.3~~$+41$:06:28 & $0.05 \pm 0.01$   & $ < 0.4$          &   0.49 \cr  
$125050.7+410738$ &     & 12:50:50.8~~$+41$:07:38 & $0.20 \pm 0.02$   & $ < 0.2$          &   2.20 \cr  
$125051.0+410742$ &     & 12:50:51.0~~$+41$:07:43 & $0.05 \pm 0.01$   & $ < 0.5$          &   0.44 \cr  
$125051.0+410646$ &     & 12:50:51.1~~$+41$:06:47 & $0.19 \pm 0.02$   & $ < 0.2$          &   2.20 \cr  
$125052.0+410716$ &     & 12:50:52.1~~$+41$:07:16 & $0.03 \pm 0.01$   & $ < 1.3$          &   0.20 \cr  
$125052.0+410654$ &     & 12:50:52.1~~$+41$:06:55 & $0.24 \pm 0.03$   & $0.47 \pm 0.08$   &   2.20 \cr  
$125052.5+410714$ &     & 12:50:52.6~~$+41$:07:15 & $0.84 \pm 0.05$   & $0.50 \pm 0.05$   &   7.40 \cr  
$125052.5+410701$ &     & 12:50:52.6~~$+41$:07:02 & $1.49 \pm 0.06$   & $ < 0.06$         &  17.00 \cr  
$125052.7+410718$ & X-3 & 12:50:52.7~~$+41$:07:19 & $4.6 \pm 0.1$     & $0.272 \pm 0.009$ &  51.00 \cr  
$125052.9+410745$ &     & 12:50:52.9~~$+41$:07:46 & $0.08 \pm 0.02$   & $ < 0.3$          &   0.78 \cr  
$125052.9+410707$ &     & 12:50:53.0~~$+41$:07:07 & $0.12 \pm 0.03$   & $ < 0.4$          &   1.10 \cr  
$125053.0+410712$ & X-2 & 12:50:53.1~~$+41$:07:13 & $5.4 \pm 0.1$     & $0.176 \pm 0.008$ &  59.00 \cr  
$125053.2+410718$ &     & 12:50:53.3~~$+41$:07:18 & $0.06 \pm 0.02$   & $ < 0.7$          &   0.41 \cr  
$125053.2+410659$ &     & 12:50:53.3~~$+41$:06:60 & $0.32 \pm 0.03$   & $0.19 \pm 0.03$   &   3.60 \cr  
$125053.3+410713$ & X-1 & 12:50:53.3~~$+41$:07:14 & $9.4 \pm 0.1$     & $0.44 \pm 0.01$   & 140.00 \cr  
$125053.3+410702$ &     & 12:50:53.3~~$+41$:07:03 & $0.11 \pm 0.02$   & $0.35 \pm 0.10$   &   1.10 \cr  
$125053.7+410718$ &     & 12:50:53.7~~$+41$:07:19 & $1.39 \pm 0.06$   & $0.37 \pm 0.02$   &  14.00 \cr  
$125053.7+410713$ &     & 12:50:53.8~~$+41$:07:13 & $0.28 \pm 0.03$   & $0.22 \pm 0.04$   &   3.00 \cr  
$125054.0+410629$ &     & 12:50:54.1~~$+41$:06:30 & $0.19 \pm 0.02$   & $ < 0.2$          &   2.20 \cr  
$125054.3+410746$ &     & 12:50:54.4~~$+41$:07:47 & $0.33 \pm 0.03$   & $0.59 \pm 0.08$   &   2.70 \cr  
$125054.5+410636$ &     & 12:50:54.6~~$+41$:06:37 & $0.03 \pm 0.01$   & $ < 0.6$          &   0.26 \cr  
$125055.0+410634$ &     & 12:50:55.0~~$+41$:06:34 & $0.05 \pm 0.01$   & $ < 0.5$          &   0.47 \cr  
$125055.2+410831$ &     & 12:50:55.2~~$+41$:08:31 & $0.04 \pm 0.01$   & $ < 0.4$          &   0.42 \cr  
$125055.6+410719$ &     & 12:50:55.6~~$+41$:07:20 & $0.28 \pm 0.03$   & $ < 0.1$          &   3.20 \cr  
$125055.7+410704$ &     & 12:50:55.7~~$+41$:07:05 & $0.04 \pm 0.01$   & $ < 0.5$          &   0.34 \cr  
$125056.3+410655$ &     & 12:50:56.3~~$+41$:06:55 & $1.05 \pm 0.05$   & $0.60 \pm 0.04$   &   8.50 \cr  
$125056.8+410705$ &     & 12:50:56.8~~$+41$:07:06 & $0.016 \pm 0.007$ & $ < 1.0$          &   0.10 \cr  
$125058.4+410813$ &     & 12:50:58.4~~$+41$:08:14 & $0.014 \pm 0.006$ & $ < 1.2$          &   0.09 \cr  
$125101.9+410855$ &     & 12:51:02.0~~$+41$:08:56 & $0.025 \pm 0.008$ & $ < 0.7$          &   0.19 \cr  
$125104.0+410738$ &     & 12:51:04.1~~$+41$:07:39 & $0.041 \pm 0.009$ & $ < 0.5$          &   0.34 \cr  
$125111.0+410850$ &     & 12:51:11.1~~$+41$:08:50 & $0.06 \pm 0.01$   & $0.4 \pm 0.1$     &   0.54 \cr  
\enddata
\end{deluxetable}

\begin{deluxetable}{lccccc}
\tablenum{7}
\tablewidth{6.3in}
\tablecolumns{6}
\tablecaption{Catalog of Discrete Sources in NGC 4579\label{sourcetable4579}}
\tablehead{
\colhead{Official Name} &
\colhead{} &
\colhead{Coordinates} &
\colhead{Count Rate} &
\colhead{} &
\colhead{$L_{\rm X}$(0.5--8~keV)} 
\cr
\colhead{(CXOU J)} &
\colhead {Alias} &
\colhead{(J2000)} &
\colhead{($10^{-2}~{\rm s}^{-1}$)} &
\colhead{Color} &
\colhead{($10^{37}~{\rm erg~s}^{-1}$)} 
}
\startdata
$123739.7+114806$ &     & 12:37:39.8~~$+11$:48:06 & $0.016 \pm 0.006$ & $ < 0.9$          &   1.60 \cr  
$123740.2+114648$ &     & 12:37:40.3~~$+11$:46:48 & $0.027 \pm 0.008$ & $ < 0.9$          &   2.80 \cr  
$123740.2+114727$ & X-1 & 12:37:40.3~~$+11$:47:27 & $3.30 \pm 0.08$   & $0.35 \pm 0.01$   & 1050.00\cr  
$123740.4+114835$ &     & 12:37:40.4~~$+11$:48:36 & $0.034 \pm 0.009$ & $0.8 \pm 0.3$     &   3.80 \cr  
$123741.9+115004$ &     & 12:37:41.9~~$+11$:50:05 & $0.07 \pm 0.01$   & $ < 0.4$          &  10.00 \cr  
$123742.3+114909$ &     & 12:37:42.3~~$+11$:49:09 & $0.04 \pm 0.01$   & $ < 0.6$          &   4.60 \cr  
$123742.6+114825$ &     & 12:37:42.6~~$+11$:48:25 & $0.024 \pm 0.008$ & $ < 0.9$          &   2.40 \cr  
$123742.7+114945$ &     & 12:37:42.8~~$+11$:49:45 & $0.04 \pm 0.01$   & $0.7 \pm 0.3$     &   5.40 \cr  
$123743.0+114759$ &     & 12:37:43.0~~$+11$:47:60 & $0.020 \pm 0.007$ & $ < 1.0$          &   1.90 \cr  
$123743.2+114901$ &     & 12:37:43.2~~$+11$:49:01 & $0.42 \pm 0.04$   & $0.27 \pm 0.04$   &  70.00 \cr  
$123743.2+114912$ &     & 12:37:43.2~~$+11$:49:13 & $0.05 \pm 0.01$   & $ < 0.7$          &   6.00 \cr  
$123743.7+114824$ &     & 12:37:43.7~~$+11$:48:25 & $0.09 \pm 0.01$   & $0.34 \pm 0.09$   &  14.00 \cr  
$123744.1+114917$ &     & 12:37:44.1~~$+11$:49:17 & $0.11 \pm 0.02$   & $0.5 \pm 0.1$     &  15.00 \cr  
$123744.3+114834$ &     & 12:37:44.3~~$+11$:48:35 & $0.032 \pm 0.009$ & $ < 0.7$          &   3.90 \cr  
$123744.5+114921$ &     & 12:37:44.5~~$+11$:49:22 & $0.08 \pm 0.02$   & $1.0 \pm 0.3$     &   8.00 \cr  
$123745.3+114818$ &     & 12:37:45.4~~$+11$:48:19 & $0.025 \pm 0.008$ & $ < 0.9$          &   2.50 \cr  
$123746.3+114958$ &     & 12:37:46.3~~$+11$:49:59 & $0.04 \pm 0.01$   & $0.7 \pm 0.3$     &   5.30 \cr  
$123746.4+115012$ &     & 12:37:46.5~~$+11$:50:13 & $0.011 \pm 0.005$ & $ < 1.1$          &   1.10 \cr  
$123746.8+114831$ &     & 12:37:46.8~~$+11$:48:32 & $0.027 \pm 0.008$ & $ < 0.9$          &   2.80 \cr  
$123747.4+115105$ &     & 12:37:47.5~~$+11$:51:06 & $0.07 \pm 0.01$   & $0.4 \pm 0.1$     &  10.00 \cr  
$123753.8+115020$ &     & 12:37:53.9~~$+11$:50:20 & $0.29 \pm 0.03$   & $0.24 \pm 0.03$   &  50.00 \cr  
\enddata
\end{deluxetable}


\begin{deluxetable}{lrrr}
\tablenum{8}
\tablewidth{4.5in}
\tablecolumns{4}
\tablecaption{Power Budget\label{budgettable}}
\tablehead{
\colhead{} &
\multicolumn{3}{c}{0.5--10 keV Luminosity (erg~s$^{-1}$)} \cr
\colhead{} &
\multicolumn{3}{c}{\hrulefill} \cr
\colhead{} &
\colhead{NGC 404} &
\colhead{NGC 4736} &
\colhead{NGC 4579} 
}
\startdata
AGN                         & \dots              & \dots               & $2.9\times 10^{41}$ \cr
Diffuse or Thermal Emission\tablenotemark{\;a} & $1\times 10^{37}$  & $6\times 10^{38}$   & $1.6\times 10^{40}$ \cr
UXBs                        & \dots              & \dots               & $1.3\times 10^{40}$ \cr
Other discrete sources      & \dots  & $3.7\times 10^{39}$ & $2\times 10^{39}$ \cr

& \multicolumn{1}{c}{\hrulefill} & 
\multicolumn{1}{c}{\hrulefill} & \multicolumn{1}{c}{\hrulefill} \cr
{\it Chandra} Total            & $1\times 10^{37}$  & $4\times 10^{39}$ & $3.2\times 10^{41}$ \cr
\sidehead{}
{\it ASCA} Total\tablenotemark{\;b} & $<5\times 10^{37}$  & $3\times 10^{39}$ & $3.0\times 10^{41}$ \cr
\tablenotetext{a\;} {This is the {\it total} luminosity of the entire diffuse source, estimated after
subtracting the contribution of point sources embedded in it.}
\tablenotetext{b\;} {References. --
NGC~404: \citet{rsw01};
NGC~4736: \citet{rwo99};
NGC~4579: \citet{t_etal98}}
\enddata
\end{deluxetable}



\clearpage

\begin{figure}
\plotone{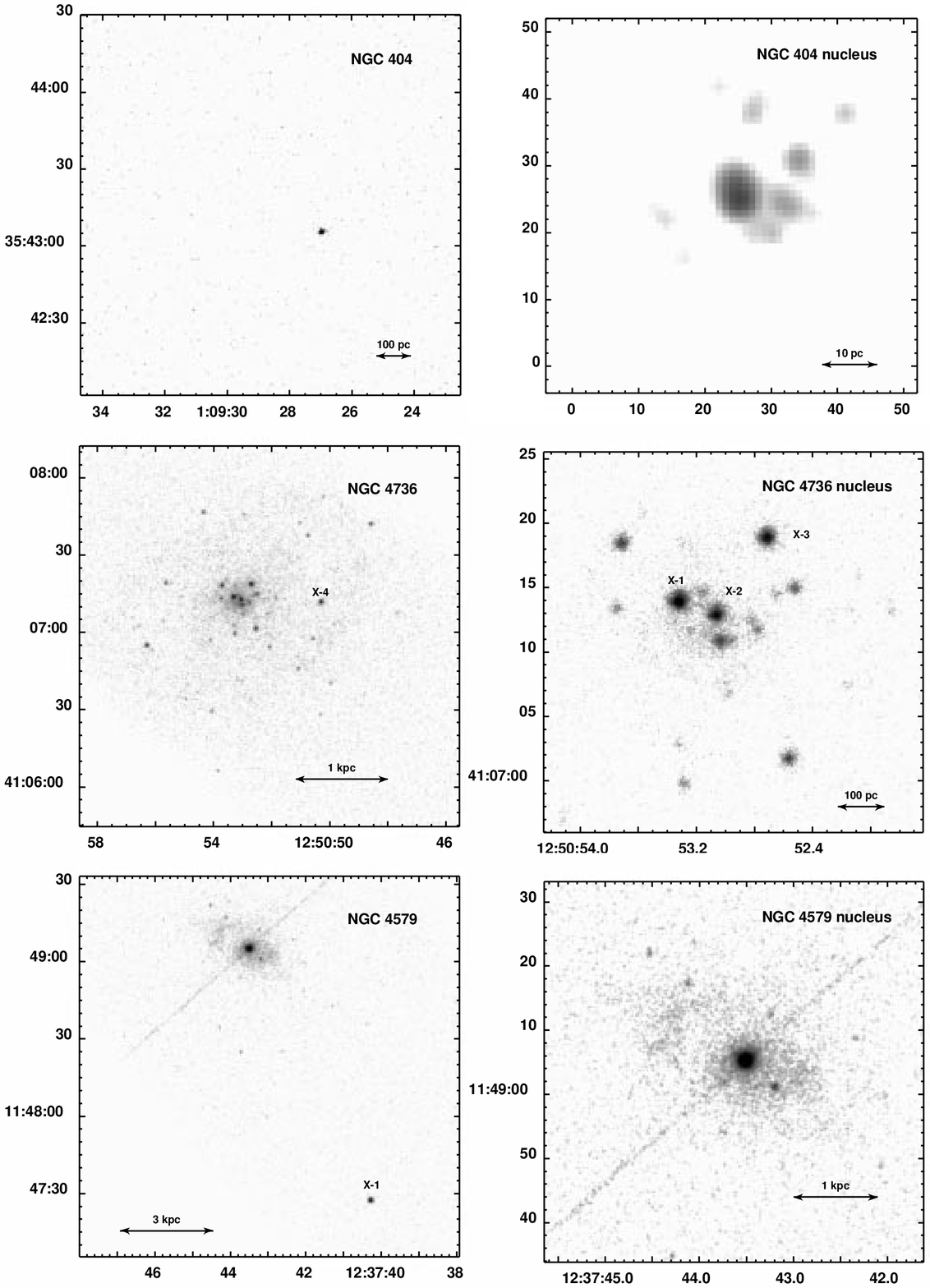}
\bigskip
\caption{\label{imgmont} Montage of \chandra\ images of the target
galaxies. For each galaxy we show an image encompassing all the
interesting features in the field of view as well as an enlarged view
of the nucleus. All but one of the frames are labeled with the right
ascension along the horizontal axis and the declination along the
vertical axis (the coordinate epoch is J2000, north is up, and east is
to the left). The image of the nucleus of NGC~404 has been deconvolved
using the Lucy-Richardson algorithm; its frame is calibrated in
pixels, where each pixel corresponds to $0.\!\!^{\prime\prime}15$.
The discrete ``blobs'' around the nucleus of NGC~404 are not
discernible in the original, unprocessed image. A horizontal bar in
each image gives the linear scale at the distance of that
galaxy. Bright X-ray sources, referred to in the text and in the
tables, are labeled on the images.}
\end{figure}

\begin{thebibliography}{}

\bibitem[\protect\citeauthoryear{Anders \& Grevesse}{1989}]{ag89} 
Anders, E. \& Grevesse, N. 1989, Geochimica et Cosmochimica Acta, 53, 197

\bibitem[\protect\citeauthoryear{Angelini, Lowenstein, \&
Mushotzky}{Angelini et al.}{2001}]{alm01} 
Angelini, L., Loewenstein, M., \& Mushotzky, R. F.  2001, ApJ, 557, L35

\bibitem[\protect\citeauthoryear{Arnaud}{1996}]{a96}  
Arnaud, K. 1996, in ASP Conf. Ser. 101, Astronomical Data Analysis Software
and Systems V, eds. G. Jacoby \& J. barnes (San Francisco: ASP), 17

\bibitem[\protect\citeauthoryear{Barret}{2001}]{barret01} 
Barret, D. 2001, in ``Proceedings of 33rd COSPAR Scientific Assembly (Warsaw,
July 2000),'' Advances in Space Research, in press (astro-ph/0101295)

\bibitem[\protect\citeauthoryear{Barth \& Shields}{2000}]{bs00} 
Barth, A. J. \& Shields, J. C. 2000, PASP, 112, 753

\bibitem[\protect\citeauthoryear{Barth, Filippenko, \& Moran}{Barth et 
al.}{1999}]{barth_etal99} 
Barth, A. J., Filippenko, A. V., \& Moran, E. C. 1999, \apj, 525, 673

\bibitem[\protect\citeauthoryear{Barth et al.}{2001}]{barth_etal01} 
Barth, A. J., Ho, L. C., Filippenko, A. V., Rix, H.-W.,
\& Sargent, W. L. W. 2001, \apj, 546, 205

\bibitem[\protect\citeauthoryear{Bauer et al.}{2001}]{bauer_etal01} 
Bauer, F. E., Brandt, W. N., Samburna, R. M., Chartas, G., Garmire, G. P.,
Kaspi, S., \& Netzer, H. 2001, \aj, in press (astro-ph/0104035)

\bibitem[\protect\citeauthoryear{Binette et al.}{1994}]{binette_etal94} 
Binette, L., Magris, C. G., Stasinska, G., \& Bruzual, A. G. 1994, A\&A, 292, 13

\bibitem[\protect\citeauthoryear{Blandford \& Begelman}{1999}]{bb99} 
Blandford, R. D., \& Begelman, M. C. 1999, \mnras, 303, L1

\bibitem[\protect\citeauthoryear{Blanton, Sarazin, \& Irwin}{Blanton 
et al.}{2001}]{bsi01}
Blanton, E. L., Sarazin, C. L., \& Irwin, J. A. 2001, \apj, 552, 106

\bibitem[\protect\citeauthoryear{B\"oker}{2000}]{b00} 
B\"oker, T. 2000, in ``Massive Stellar Clusters'', eds. A. Lan\c{c}on \& C. 
Boily, (San Fransisco: ASP), 227

\bibitem[\protect\citeauthoryear{Bower et al.}{1996}]{bower_etal96} 
Bower, G. A., Wilson, A. S., Heckman, T. M., \& Richstone, D. O. 1996,
\aj, 111, 1901

\bibitem[\protect\citeauthoryear{Broos et al.}{2000}]{broos_etal00} 
Broos, P., et al. 2000, User's Guide for the \verb+TARA+ Package: Document
Revision 5.8 (University Park: Penn State University),
\verb+http://www.astro.psu.edu/xray/docs+

\bibitem[\protect\citeauthoryear{Carollo}{1999}]{c99} 
Carollo, M. 1999, \apj, 523, 566

\bibitem[\protect\citeauthoryear{Cervi\~no, Mas-Hesse, \& Kunth}{Cervi\~no et 
al.}{1997}]{cmk97}
Cervi\~no, M., Mas-Hesse, J. M. \& Kunth, D. 1997, RMxAA (Conf. Ser.), 6,188

\bibitem[\protect\citeauthoryear{Chu \& Mac Low}{1990}]{cm90} 
Chu, Y.-H., \& Mac Low, M.-M. 1990, \apj, 365, 510

\bibitem[\protect\citeauthoryear{Chartas et al.}{2000}]{chartas00} 
Chartas, G. et al. 2000, \apj, 542, 655

\bibitem[\protect\citeauthoryear{Cui, Feldkhun, \& Braun} {Cui et 
al.}{1997}]{cui97}
Cui, W., Feldkhun, D., \& Braun, R. 1997, \apj, 477, 693

\bibitem[\protect\citeauthoryear{Dalton \& Sarazin}{1995}]{dalton95} 
Dalton, W. W. \& Sarazin, C. L. 1995, \apj, 440, 280

\bibitem[\protect\citeauthoryear{de Vaucouleurs et al.}{1991}]{RC3} 
de Vaucouleurs, G., de Vaucouleurs, A., Corwin, H. G., Jr., Buta, R. J.,
Paturel, G., \& Fouqu\'e, R. 1991, Third Reference Catalogue of Bright
Galaxies (New York: Springer)

\bibitem[\protect\citeauthoryear{Dopita \& Sutherland}{1995}]{ds95} 
Dopita, M. A. \& Sutherland, R. S. 1995, \apj, 455, 468

\bibitem[\protect\citeauthoryear{Dopita \& Sutherland}{1996}]{d_etal96} 
Dopita, M. A. et al. in ``The Physics of LINERs in View of Recent
Observations'', eds. M. Eracleous et al. (San Francisco: ASP), 44

\bibitem[\protect\citeauthoryear{Dressler}{1987}]{dressler87} 
Dressler, A. 1987, \apj, 317, 1

\bibitem[\protect\citeauthoryear{Eracleous \& Halpern}{1994}]{eh94} 
Eracleous, M. \& Halpern, J. P. 1994, \apjs, 90, 1

\bibitem[\protect\citeauthoryear{Eracleous \& Halpern}{2001}]{eh01} 
Eracleous, M. \& Halpern, J. P. 2001, \apj, 554, 240

\bibitem[\protect\citeauthoryear{Eracleous, Livio, \& Binette}{Eracleous 
et al.}{1995}]{elb95}
Eracleous, M., Livio, M., \& Binette, L. 1995, \apj, 445, L1

\bibitem[\protect\citeauthoryear{Eracleous, Patterson, \& 
Halpern}{Eracleous et al.}{1991}]{eph91}
Eracleous, M., Patterson, J., \& Halpern, J. P. 1991, \apj, 370, 330

\bibitem[\protect\citeauthoryear{Eracleous, Sambruna, \& 
Mushotzky}{Eracleous et al.}{2000}]{esm00}
Eracleous, M., Sambruna, R. M., \& Mushotzky, R. F. \apj, 537, 654 

\bibitem[\protect\citeauthoryear{Fabbiano, Zezas, \& Murray}{Fabbiano
et al.}{2001}]{fzm01}
Fabbiano, G., Zezas, A., \& Murray, S. S. 2001, \apj, in press
(astro-ph/0102256)

\bibitem[\protect\citeauthoryear{Faber et al.}{1989}]{f_etal89} 
Faber, S. M., Wegner, G., Burstein, D., Davies, R. L., Dressler, A.,
Lynden-Bell, D., \& Terlevich, R. J. 1989, \apjs, 69, 763

\bibitem[\protect\citeauthoryear{Falcke et al.}{2000}]{f_etal00}  
Falcke, H., Nagar, N. M., Wilson, A. S., \& Ulvestad, J. S. 2000,
\apj, 542, 197

\bibitem[\protect\citeauthoryear{Ferland \& Netzer}{1983}]{fn83} 
Ferland, G. J. \& Netzer, H. 1983, \apj, 264, 105


\bibitem[\protect\citeauthoryear{Ferrarese et al.}{2001}]{f_etal01} 
Ferrarese, L., Pogge, R. W., Peterson, B. M., Merrit, D., Wandel, A. \& Joseph,
C. L. 2001, \apjl, submitted (astro-ph/0104380)

\bibitem[\protect\citeauthoryear{Filippenko}{1996}]{f96} 
Filippenko, A. V. 1996, in ``The Physics of LINERs in View of Recent
Observations'', eds. M. Eracleous et al. (San Francisco: ASP), 17

\bibitem[\protect\citeauthoryear{Filippenko \& Terlevich}{1992}]{ft92} 
Filippenko, A. V. \& Terlevich, R. 1992, \apj, 397, L79

\bibitem[\protect\citeauthoryear{Freeman et al.}{2001}]{freeman01} 
Freeman, P. E., Kashyap, V., Rosner, R., \& Lamb, D. Q. 2001, \apj, submitted

\bibitem[\protect\citeauthoryear{Gebhardt et al.}{2000}]{gebhardt00} 
Gebhardt, K. et al. 2000, \apj, 539, L3

\bibitem[\protect\citeauthoryear{Griffiths et al.}{2000}]{g_etal00} 
Griffiths, R. E., Ptak, A. F., Feigelson, E. D., Garmire, G., Townsley, L.,
Brandt, W. N., Sambruna, R. M., \& Bregman, J. N. 2000, Science, 290, 1325

\bibitem[\protect\citeauthoryear{Halderson et al.}{2001}]{h_etal01} 
Halderson, E. L., Moran, E. C., Filippenko, A. V., \& Ho., L. C. 2001, \aj,
in press (astro-ph/0105069).

\bibitem[\protect\citeauthoryear{Halpern \& Steiner}{1983}]{hs83} 
Halpern, J. P. \& Steiner, J. E. 1983, \apj, 269, L37

\bibitem[\protect\citeauthoryear{Heckman}{1980}]{h80} 
Heckman, T. M 1980, A\&A, 87, 152

\bibitem[\protect\citeauthoryear{Helfand \& Moran}{2001}]{hm01} 
Helfand, D. J. \& Moran, E. C. 2001, \apj, 554, 27

\bibitem[\protect\citeauthoryear{Hempellmann et al.}{1995}]{h_etal95} 
Hempellmann, A., Schmidt, J. H. M. M., Schultz, M., R\"udiger, G., \&
Stepien, K. 1995, \aa, 294, 515

\bibitem[\protect\citeauthoryear{Ho}{1999}]{h99} 
Ho, L. C. 1999, \apj, 516, 672

\bibitem[\protect\citeauthoryear{Ho et al.}{2001}]{ho_etal01} 
Ho, L. C. et al. 2001, \apj, 549, L51

\bibitem[\protect\citeauthoryear{Ho et al.}{1997a}]{hfs_a97} 
Ho, L. C., Filippenko, A. V., \& Sargent, W. L. W. 1997a, \apjs, 112, 315

\bibitem[\protect\citeauthoryear{Ho et al.}{1997b}]{hfs_b97} 
Ho, L. C., Filippenko, A. V., Sargent, W. L. W., \& Peng, C. 1997b, \apjs, 112,
391

\bibitem[\protect\citeauthoryear{Ho et al.}{1997c}]{hfs_c97} 
Ho, L. C., Filippenko, A. V., \& Sargent, W. L. W. 1997c, \apj, 487, 568

\bibitem[\protect\citeauthoryear{Ho et al.}{2000}]{h_etal00} 
Ho, L., Rudnick, G,. Rix, H.-W., Shields, J. C.,  McIntosh, D. H., Filippenko,
A. V., Sargent, W. L. W., \& Eracleous, M. 2000, \apj, 000, 000

\bibitem[\protect\citeauthoryear{Hornschemeier et al.}{2001}]{annh01} 
Hornschemeier, A. E. et al. 2001, \apj, in press (astro-ph/0101494)

\bibitem[\protect\citeauthoryear{Kaastra}{1992}]{kaastra92} 
Kaastra, J .S. 1992, An X-Ray Spectral Code for Optically Thin Plasmas
(Internal SRON-Leiden Report, updated version 2.0)

\bibitem[\protect\citeauthoryear{King et al.}{2001}]{king01} 
King, A. R., Davies, M. B., Ward, M. J., Fabbiano, G., \& Elvis, M.
2001, \apj, 552, L109

\bibitem[\protect\citeauthoryear{Komossa, B\"ohringer, \& Huchra}{Komossa 
et al.}{1999}]{kbc99}
Komossa, S., B\"ohringer, H., \& Huchra, J. P. 1999, A\&A, 349, 88

\bibitem[\protect\citeauthoryear{Lasota et al.}{1996}]{l_etal96} 
Lasota, J. P., et al 1996, \apj, 462, 142

\bibitem[\protect\citeauthoryear{Liedahl, Osterheld, \& Goldstein}{Liedahl et 
al}{1995}]{liedahl95}
Liedahl, D. A., Osterheld, A. l., \& Goldstein, W. H. 1995, \apjl, 438, 115

\bibitem[\protect\citeauthoryear{Lira et al.}{2001}]{lira01}
Lira, P., Ward, M., Zezas, A., Alonso-Herrero, A. \& Ueno , S. 2001,
\mnras, in press (astro-ph/0109198)

\bibitem[\protect\citeauthoryear{Mc~Elroy}{1995}]{mcelroy95} 
Mc~Elroy, D. B. 1995, \apjs, 100, 105

\bibitem[\protect\citeauthoryear{Maoz et al.}{1995}]{m_etal95} 
Maoz, D., Filippenko, A. V., Ho, L. C., Rix, H.-W., Bahcall, J. N., Schneider,
D. P., \& Macchetto, D. 1995, \apj, 440, 91

\bibitem[\protect\citeauthoryear{Maoz et al.}{1996}]{m_etal96} 
Maoz, D., Filippenko, A. V., Ho, L. C., Macchetto, D., Rix, H.-W., \& 
Schneider, D. P.,1996, \apjs, 107, 215

\bibitem[\protect\citeauthoryear{Maoz et al.}{1998}]{m_etal98} 
Maoz, D., Koratkar, A., Shields, J. C., Ho, L. C., Filippenko, A. V., \&
Sternberg, A. 1998, \aj, 116, 55

\bibitem[\protect\citeauthoryear{Merrit \& Ferrarese}{2001}]{mf01} 
Merrit, D. \& Ferrarese, L. 2001, \apj, 547, 140

\bibitem[\protect\citeauthoryear{Mewe, Gronenschild, \& van den Oord} 
{Mewe et al.}{1985}]{mewe85} 
Mewe, R., Gronenschild, E. H. B. M., \& van den Oord, G. H. J. 1985,
A\&AS, 62, 197

\bibitem[\protect\citeauthoryear{Mewe, Lemen, \& van den 
Oord}{Mewe et al.}{1986}]{mewe86} 
Mewe, R., Lemen, J. R., \& van den Oord, G. H. J. 1986, A\&AS, 65, 511

\bibitem[\protect\citeauthoryear{Mitsuda et al.}{1984}]{mitsuda84} 
Mitsuda. K. et al. 1984, PASJ, 36, 741

\bibitem[\protect\citeauthoryear{Mitsuda et al.}{1989}]{mitsuda89} 
Mitsuda, K. H., Inoue, H., Nakamura, N., \& Tanaka, Y. 1989, PASJ, 41, 97

\bibitem[\protect\citeauthoryear{Monet et al.}{1996}]{monet96} 
Modet, D. et al. 1996, USNO-SA2.0, U.S. Naval Observatory,
Washington, D.C.

\bibitem[\protect\citeauthoryear{Morrison \& McCammon}{1983}]{mm83}  
Morrison, R. \& McCammon, D. 1983, ApJ, 270, 119

\bibitem[\protect\citeauthoryear{Mushotzky et al.}{2000}]{m_etal00} 
Mushotzky, R., Angelini, L., Arnaud, K., \& Lowenstein, M. 2000, HEAD, 32, 2102

\bibitem[\protect\citeauthoryear{Nandra et al.}{1997a}]{nandra97a} 
Nandra, K., George, I. M., Mushotzky, R. F., Turner, T. J., \&
Yaqoob, T. 1997, \apj, 476, 70



\bibitem[\protect\citeauthoryear{Nagase}{1989}]{nagase89} 
Nagase, F. 1989, PASJ, 41, 1

\bibitem[\protect\citeauthoryear{Narayan \& Yi}{1994}]{ny94} 
Narayan, R. \& Yi, I. 1994, \apj, 428, L13

\bibitem[\protect\citeauthoryear{Narayan \& Yi}{1995}]{ny95} 
Narayan, R. \& Yi, I. 1995, \apj, 444, 231

\bibitem[\protect\citeauthoryear{Narayan, Yi, \& Mahadevan}{Narayan et 
al.}{1996}]{nym96} 
Narayan, R., Yi, I., \& Mahadevan, R. 1996, A\&AS, 120, C287

\bibitem[\protect\citeauthoryear{Pogge}{1989}]{p89} 
Pogge, R. W. 1989, \apjs, 71, 433

\bibitem[\protect\citeauthoryear{Pogge et al.}{2000}]{p_etal00} 
Pogge, R. W., Maoz, D., Ho, L. C., \& Eracleous, M. 2000, ApJ, 532, 323 

\bibitem[\protect\citeauthoryear{Primini, Forman, \& Jones}{Primini et 
al.}{1993}]{pfj93}
Primini, F. A., Forman, W., \& Jones, C. 1993, \apj, 410, 615

\bibitem[\protect\citeauthoryear{Ptak et al.}{1998}]{p_etal98}  
Ptak, A., Yaqoob, T., Mushotzky, R., Serlemitsos, P., \& Griffiths,
R. 1998, \apj, 501, L37

\bibitem[\protect\citeauthoryear{Ptak et al.}{1999}]{p_etal99} 
Ptak, A., Serlemitsos, P. J., Mushotzky, R. F., \& Yaqoob, T. 1999, \apjs
120, 179

\bibitem[\protect\citeauthoryear{Roberts, Warwick, \& Ohashi}{Roberts et 
al.}{1999}]{rwo99}
Roberts, T. P.,  Warwick, R. S., \& Ohashi, T. 1999, \mnras, 304, 52

\bibitem[\protect\citeauthoryear{Roberts, Schurch, \& Warwick}{Roberts et 
al.}{2001}]{rsw01}
Roberts, T. P., Schurch, N. J., \& Warwick, R. S., 2001, \mnras, 324, 737

\bibitem[\protect\citeauthoryear{Sarazin, Irwin, \& Bregman}{Sarazin et 
al.}{2001}]{sib01}
Sarazin, C. L., Irwin, J. A., \& Bregman, J. N. 2001, \apj, in press
(astro-ph/0104070)

\bibitem[\protect\citeauthoryear{Schlegel, Finkbeiner, \& 
Davis}{Schlegel et al.}{1998}]{sfd98}
Schlegel, D. J., Finkbeiner, D. P., Davis, M. 1998, \apj, 500, 525

\bibitem[\protect\citeauthoryear{Shields}{1992}]{s92} 
Shields, J. C. 1992, \apj, 399, L27

\bibitem[\protect\citeauthoryear{Shields et al.}{2000}]{s_etal00} 
Shields, J. C., Rix, H.-W., McIntosh, D. H., Ho, L. C., Rudnick, G.,
Filippenko, A. V., Sargent, W. L. W., \& Sarzi, M. 2000, \apj, 534, L27

\bibitem[\protect\citeauthoryear{Sipior \& Eracleous}{2000}]{se00} 
Sipior, M., \& Eracleous, M. 2000, HEAD, 32, 1507

\bibitem[\protect\citeauthoryear{Smith et al.}{1994}]{s_etal94} 
Smith, B. J., Harvey, P. M., Colom\'e, C., Zhang, C. Y., DiFrancesco, J.,
\& Pogge, R. W. 1994, \apj, 425, 91

\bibitem[\protect\citeauthoryear{Stark et al.}{1992}]{stark92} 
Stark, A. A., Gammie, C. F., Wilson, R. W., Bally, J.,  Linke,
R. A., Heiles, C., \& Hurwitz, M. 1992, \apjs, 79, 77

\bibitem[\protect\citeauthoryear{Taniguchi et al.}{1996}]{tani96} 
Taniguchi, Y., Ohyama, Y., Yamada, T., Mouri, H., \& Yoshida, M.
1996, \apj, 467, 215.

\bibitem[\protect\citeauthoryear{Tennant et al.}{2001}]{tennant01} 
Tennant, A. F., Wu, K., Ghosh, K. K., Kolodziejczak, J. J., \&
Schwartz, D. A. 2001, \apj, 549, L43

\bibitem[\protect\citeauthoryear{Terashima et al.}{1998}]{t_etal98} 
Terashima, Y., Kunieda, H., Misaki, K., Mushotzky, R. F., Ptak, A. F., \&
Reichert, G. A. 1998, \apj, 503, 212

\bibitem[\protect\citeauthoryear{Terashima et al.}{2000a}]{t_etal_a00} 
Terashima, Y., Ho, L. C., Ptak, A. F., Mushotzky, R. F., Serlemitsos, P. J., \&
Yaqoob, T. 2000b, \apj, 533, 729


\bibitem[\protect\citeauthoryear{Terashima et al.}{2000c}]{t_etal_b00} 
Terashima, Y., Ho, L. C., Ptak, A. F., Yaqoob, T., Kunieda, H., Misaki, K.,
\& Serlemitsos, P. J. 2000c, \apj, 535, L79

\bibitem[\protect\citeauthoryear{Terlevich \& Melnick}{1985}]{tm85} 
Terlevich, R. \& Melnick, J. 1985, \mnras, 213, 841

\bibitem[\protect\citeauthoryear{Titarchuk}{1994}]{tit94} 
Titarchuk, L. 1994, \apj, 434, 450

\bibitem[\protect\citeauthoryear{Tozzi et al}{2001}]{t_etal01} 
Tozzi, P. et al. 2001, \apj, submitted (astro-ph/0103198)

\bibitem[\protect\citeauthoryear{Tully}{1988}]{tully88} 
Tully, R. B. 1998, Nearby Galaxies Catalog (Cambridge: Cambridge Univ. Press)

\bibitem[\protect\citeauthoryear{Turner \& Ho}{1994}]{th94} 
Turner, J. L. \& Ho, P. T. P 1994, \apj, 421, 122

\bibitem[\protect\citeauthoryear{Van~Bever \& Vanbeveren}{2000}]{vb00} 
Van~Bever, J. \& Vanbeveren, D. 2000, A\&A, 358, 462

\bibitem[\protect\citeauthoryear{Wang et al.}{1991}]{whhw91} 
Wang, Q., Hamilton, T., Helfand, D. J., \& Wu, X. 1991, \apj, 374, 475

\bibitem[\protect\citeauthoryear{Wang \& Helfand}{1991}]{wh91} 
Wang, Q. \& Helfand, D. J. 1991, \apj, 370, 541

\bibitem[\protect\citeauthoryear{Ward et al.}{2000}]{w_etal00} 
Ward, M. J., Zezas, A., Lira, P., Prestwich, A., Murray, S., Alonso-Herrero,
A., \& Ueno, S. 2000, HEAD, 32, 1805

\bibitem[\protect\citeauthoryear{White, Stella, \& Parmar}{White et 
al.}{1988}]{wsp88}
White, N. E., Stella, L., \& Parmar, A. N. 1988, \apj, 324, 363

\bibitem[\protect\citeauthoryear{Storchi-Bergmann, Baldwin, \& 
Wilson}{Storchi-Bergmann et al.}{1993}]{sbw93} 
Storchi-Bergmann, T., Baldwin, J. A., \& Wilson, A. S. 1993, \apj, 410, L11

\bibitem[\protect\citeauthoryear{Woo, Clark, \& Levine}{Woo et 
al.}{1995}]{woo95}
Woo, J. W., Clark, G. W., \& Levine, A. 1995, \apj, 449, 880






\end{thebibliography}
\end{document}